\documentclass{emulateapj}
\usepackage{color}
\usepackage{natbib}
\bibpunct{(}{)}{;}{a}{}{,} %to follow the A&A style
\usepackage{graphicx}
\usepackage{txfonts}
\def\kms{km s$^{-1}$}
\def\kmss{km s$^{-1}$\space}

\def\microns{$\mu$m\space}
\def\arcsec{$^{\prime\prime}$}
\def\arcsecs{$^{\prime\prime}$\space}

\def\arcmin{$^{\prime}$}

\def\deg{$^{\circ}$}
\def\degs{$^{\circ}$\space}

\def\h2{H$_2$}
\def\n2h{N$_2$H$^+$}

\def\cuv{C\,{\sc ii}}
\def\cii{[C\,{\sc ii}]\space}
\def\ciis{[C\,{\sc ii}]}
\def\cis{C\,{\sc i}}
\def\ci{C\,{\sc i}\space}

\def\oi{O\,{\sc i}\space}

\def\ki{K\,{\sc i}\space}
\def\hi{H\,{\sc i}\space}
\def\his{H\,{\sc i}}
\def\hii{H\,{\sc ii}\space}
\def\13co{$^{13}$CO}
\def\c18o{C$^{18}$O}
\def\co{$^{12}$CO\space}
\def\cos{$^{12}$CO}

\def\c+{C$^+$}

\def\h2{H$_2$}
\usepackage{amstext}
\shorttitle{ Thermal Pressure from \cii emission and \h2 UV absorption}
\shortauthors{Velusamy et al.}
\begin{document}
   \title{Thermal Pressure in Diffuse \h2 Gas Measured by\\ \textit{Herschel} \cii Emission and \textit{FUSE} UV \h2 Absorption}
%\titlerunning{Thermal Pressure from \cii emission and \h2 UV absorption}
%\authorrunning{Velusamy,  Langer, Goldsmith, Pineda }
\author{T. Velusamy\altaffilmark{1}, W. D. Langer\altaffilmark{1},  P. F. Goldsmith\altaffilmark{1}, J. L. Pineda\altaffilmark{1}  }
\altaffiltext{1}{Jet Propulsion Laboratory, California Institute of
Technology, 4800 Oak Grove Drive, Pasadena, CA 91109;
Thangasamy.Velusamy@jpl.nasa.gov,  William.D.Langer@jpl.nasa.gov,  	Paul.F.Goldsmith@jpl.nasa.gov,  	Jorge.Pineda@jpl.nasa.gov }

%   \date{Received TBD; accepted TBD}

\begin{abstract}
UV absorption studies with the \textit{FUSE} satellite have made important observations of  \h2 molecular gas in Galactic interstellar translucent and diffuse clouds. Observations of the  158 \microns \cii fine structure line  with \textit{Herschel}  trace the same \h2 molecular gas   in emission. We present \cii observations along  27 lines of sight (LOSs) towards target stars of which 25    have  \textit{FUSE} \h2 UV absorption. Two stars have only  HST STIS \cuv $\lambda$2325  absorption data.     We detect \cii 158 \microns emission features in all but one  target LOS.   For three target LOSs which are close to the Galactic plane, ${|\it b|} <$ 1\deg, we also present position-velocity maps of \cii emission  observed by \textit{Herschel} HIFI in on-the-fly spectral line mapping.
We use the  velocity resolved \cii spectra observed by the HIFI instrument towards the target  LOSs observed by \textit{FUSE} to identify \cii  velocity components associated with the \h2  clouds. We analyze the observed  velocity integrated   \cii spectral line intensities  in terms of the densities and thermal pressures in the \h2 gas using the \h2 column densities and temperatures measured by the UV absorption data.   We present the \h2 gas densities and thermal pressures for 26 target LOSs and  from the \cii intensities   derive a mean thermal pressure  in the range of  $\sim$ 6100 to 7700 K cm$^{-3}$  in diffuse \h2 clouds.    We discuss the thermal pressures and densities towards 14 targets,  comparing them to results obtained using the UV absorption data for two other tracers \ci and CO.
Our results demonstrate  the richness of the far-IR \cii spectral data which is a valuable complement to  the UV \h2 absorption data for studying diffuse \h2 molecular clouds. While the UV absorption is restricted to  the directions of the target star, far-IR \cii line emission offers an   opportunity to employ velocity resolved spectral line mapping capability to study in detail the clouds' spatial and velocity structures.
\end{abstract}

\pagebreak
% \maketitle
%
%________________________________________________________________
%\vspace{-1cm}
\section{Introduction}
The evolution of interstellar clouds can be broadly classified into four different evolutionary stages as diffuse atomic, diffuse molecular, translucent, and dense molecular clouds. The diffuse molecular and translucent clouds represent stages where the gas is primarily molecular hydrogen \citep{Snow2006} and UV radiation is present throughout the cloud.
 However, these diffuse and translucent  \h2 clouds have been difficult to study throughout the Galaxy because \h2 does not emit under typical cloud temperatures, and CO, an important surrogate for H$_2$ in regions of higher extinction, is virtually absent due to UV photodissociation.   Instead these clouds have primarily been studied locally (within $\sim$1 kpc) by visual and UV absorption lines against a bright background source.
It has been proposed that Giant Molecular Cloud (GMC) formation, and thus the star formation rate in galaxies, is regulated by thermal pressure in the interstellar medium, or ISM  \citep{Cox2005,Blitz2006}.
Cloud formation over large scales depends on the thermal and dynamical state of the interstellar gas which  in turn is modulated by heating and cooling rates, the gravitational potential, and turbulent and thermal  pressures. Thus to more fully understand
the thermodynamic state of the Galactic ISM and its relation to star formation, it is important to evaluate the pressure over the full extent of the disk.  At present, the variation of thermal pressure throughout the Milky Way is
observationally poorly defined. \cite{Gerin2015} have used the 158 \microns \cii absorption spectra observed by HIFI towards a few selected  Galactic objects to derive the thermal pressure in the diffuse ISM.   In the solar neighborhood ($\sim$1 kpc) measurements of the thermal gas pressure have been made with \ci UV absorption spectra towards a sample of about 89 selected target stars \citep{Jenkins1983,Jenkins2001,Jenkins2011} and CO UV absorption towards 76 stars \citep{Goldsmith2013}.

In this paper we develop a different approach to measuring the thermal pressure that relies on emission from \ciis, the fine structure emission line of ionized carbon at 158 \microns (1.9 THz).  To test this approach we observed spectrally resolved \cii with the Heterodyne Instrument in the Far Infrared (HIFI, \cite{deGraauw2010}) on the {\it Herschel Space Observatory}\footnote{{\it Herschel} Space Observbatory is an ESA space observatory with science instruments provided by European-led Principal Investigator consortia and with important participation from NASA.} \citep{Pilbratt2010}  towards a sample of 25  lines of sight (LOS) with \h2 column densities previously derived from Far Ultraviolet Spectroscopic Explorer ({\it FUSE}) UV absorption observations \citep{Rachford2002,Sheffer2008} and two from those observed in \cuv $\lambda$2325 absorption by HST STIS \citep{Sofia2004}.  The advantage of \cii is that it is observed in emission and so can probe the thermal pressure across the Milky Way. The large--scale {\it Herschel} \cii survey of the Milky Way (GOT C+; see \cite{Langer2010}) has shown that \cii emission can be apportioned among the different phases of the interstellar medium \citep{Langer2010,Langer2014_II,Pineda2010,Pineda2013,Velusamy2010,Velusamy2014} and that a significant fraction is attributed to the emission from diffuse and translucent \h2 clouds.

 \begin{table*}[!t]
% \renewcommand{\arraystretch}{0.8}
%\centering
\begin{center}
\caption{Target lines of sight (LOS) observed by {\it Herschel} HIFI in \cii emission at 158 $\mu$$m$  }
%\vspace{-0.25cm}
\renewcommand{\tabcolsep}{0.25cm}
%\resizebox{5.0cm}{!} {
\begin{tabular}  {l l c c c c c c c c c}
\hline\hline
\multicolumn{2}{c} {Target Star}	&\multicolumn{2}{c}{	log $N$(H$_2$)}	& \multicolumn{2}{c} { log $N$(\his)}	&\multicolumn{2}{c} { log $N$(\c+)}& 	T$_{01}$(H$_2$)	&	 Distance	\\
& &	  cm$^{-2}$	& ref	&10$^{20}$ cm$^{-2}$	&ref	& 10$^{17}$ cm$^{-2}$	&ref	&(K)	&	(pc)	\\
\hline
BD +31 643 	&	G	160.491	-17.802	&	21.09	$\pm$	0.19	&	1	&	21.38	&	1	&	--	    &	 	&	73$\pm$48	&	150	\\
HD 24534  	&	G	163.081	-17.136	&	20.92	$\pm$	0.04	&	1	&	20.73	&	1	&	17.49	&	7	&	57$\pm$4	&	2100	\\
HD 34078 	&	G	172.081	-02.259	&	20.88			        &	2	&	21.19	&	5	&	---	    &		&	75	&	450	\\
HD 37021	&	G	209.006	-19.384	&	---			           &	 	&	21.68	&	4	&	17.64	&	3,8	&	70$^{11}$	&	560	\\
HD 37061	&	G	208.925	-19.274	&	---			           &	 	&	21.73	&	4	&	17.72	&	3,8	&	70$^{11}$	&	640	\\
HD 37903	&	G	206.851	-16.537	&	20.92	$\pm$	0.06	&	2,10&	21.16	&	4	&	18.02	&3,8,9	&	68$\pm$7	&	830	\\
HD 62542 	&	G	255.915	-09.237	&	20.81	$\pm$	0.21	&	1	&	20.93	&	1	&	---	     &		&	43$\pm$11	&	310	\\
HD 73882 	&	G	260.182	+00.643	&	21.11	$\pm$	0.08	&	1	&	21.11	&	1	&	---	     &		&	51$\pm$6	&	450	\\
HD 94454  	&	G	295.692	-14.726	&	20.76			       &	2	&	0.00    &	5   &	17.50	&	7	&	74	&	300	\\
HD 96675  	&	G	296.616	-14.569	&	20.82	$\pm$	0.05	&	2	&	20.66	&	1	&	---	    &		&	61$\pm$7	&	160	\\
HD 115071 	&	G	305.764	+00.152	&	20.69			       &	2	&	20.91    &	6	&	17.87	&	7	&	71	&	2700	\\
HD 144965	&	G	339.043	+08.417	&	20.79			        &	2	&	20.52  &	5	&	17.52	&	7	&	70	&	510	\\
HD 147683	&	G	344.857	+10.089	&	20.74			        &	2	&	20.72  &	5	&	17.64	&	7	&	58	&	370	\\
HD 147888	&	G	353.647	+17.709	&	20.47	$\pm$	0.05	&	2,10&	21.71	&	4	&	17.75	&3,8,9	&	44$\pm$4	&	120	\\
HD 152590	&	G	344.840	+01.830	&	20.51			        &	2	&	21.37  &	4	&	17.80	&3,8,9	&	64	&	3600	\\
HD 154368 	&	G	349.970	+03.215	&	21.46	$\pm$	0.07	&	1	&	21.00	&	1	&	---	    &		&	51$\pm$8	&	910	\\
HD 167971 	&	G	018.251	+01.684	&	20.85	$\pm$	0.12	&	1	&	21.60	&	1	&	---	    &		&	64$\pm$17	&	660	\\
HD 168076 	&	G	016.937	+00.838	&	20.68	$\pm$ 0.08	      &	1	&	20.64	&	6	&	---	   &		&	68$\pm$13	&	2000	\\
HD 170740 	&	G	021.057	-00.526	&	20.86	$\pm$	0.08	&	1	&	20.96	&	6	&	---	    &		&	70$\pm$13	&	330	\\
HD 185418 	&	G	053.602	-02.171	&	20.76	$\pm$	0.05	&	1	&	21.19	&	4	&	17.64	&	9	&	105$\pm$6	&	1200	\\
HD 192639	&	G	074.901	+01.480	&	20.75	$\pm$	0.09 	&	1,2	&	21.29	&	4	&	17.59	&	9	&	98$\pm$15	&	2100	\\
HD 200775	&	G	104.062	+14.193	&	21.15			        &	2	&	20.68	&	5	&	---	     &		&	44	&	430	\\
HD 203938 	&	G	090.558	-02.234	&	21.00	$\pm$	0.06	&	1	&	21.48	&	1	&	--- 	&		&	74$\pm$9	&	1040	\\
HD 206267 	&	G	099.290	+03.738	&	20.86	$\pm$	0.04	&	1	&	21.30	&	1	&	17.85	&	7	&	65$\pm$5	&	860	\\
HD 207198 	&	G	103.136	+06.995	&	20.83	$\pm$	0.04	&	1	&	21.14	&	1	&	17.51	& 3,8,9 &	66$\pm$5	&	1300	\\
HD 207538 	&	G	101.599	+04.673	&	20.91	$\pm$	0.06	&	1	&	21.32	&	1	&	---	    &		&	73$\pm$8	&	2000	\\
HD 210839 	&	G	103.828	+02.611	&	20.84	$\pm$	0.04	&	1	&	21.15	&	1	&	17.79	&	7	&	72$\pm$6	&	1100	\\
\hline
\end{tabular}\\
% }
\end{center}
\vspace{-0.25cm}
\noindent References: (1) \cite{Rachford2002,Rachford2009}); (2) \cite{Sheffer2008}; (3) \cite{Sofia2004}; (4) \cite{Cartledge2004}; (5) This paper: N(HI)= 5.8$\times$10$^{21}$$E(B-V)$ -2N(H$_2$); (6) This paper: from the  \hi 21 cm  $(l-V$) maps; (7) \cite{Jenkins2011}; (8) \cite{Sofia2011}; (9) \cite{Parvathi2012}; (10) \cite{Rachford2009};  (11) Assumed value. \\
\label{tab:Table1}
\end{table*}

   By combining our HIFI \cii emission spectra with {\it FUSE} and STIS direct detections of \h2 or \c+  in absorption  we can better constrain many of the physical conditions in the clouds including the density and pressure of the  \hi and  \h2 gas components.  The availability of the thermal pressure and density in the literature from the \ci and CO data on a subset of our \cii sample allows us to verify our approach.    This validation is a step forward in the interpretation of \cii  emission  % , and by extension in external Galaxies,
    as a viable alternative to \h2 absorption studies of diffuse molecular clouds in the Galaxy.

An outline of our paper is as follows. In Section~\ref{sec:observations} we present the pointed observations towards the target selections and  on-the-fly (OTF) spectral line mapping  along three LOSs, and describe the data reduction.  In Section~\ref{sec:results} we present the \cii  spectra for all 27 LOSs and spatial-velocity maps towards three targets.   For the three mapped LOSs we compare the  distributions of \cii with publicly available  \hi  and \co or $^{13}$CO  data.  We analyze and interpret the \cii intensities to derive the thermal pressures and densities using  \h2 column densities measured in the UV absorption. In Section~\ref{sec:discussion} we present a comparison of the thermal pressures derived by \cii data with those available from  the \ci survey.
We summarize our results in Section~\ref{sec:summary}.

\section{Observations}
\label{sec:observations}

Our sample consists of the 25 {\it FUSE} targets  with the highest \h2 column densities and two (HD 37021 \& HD 37061) from the \cuv $\lambda$2325 sample for which there are no {\it FUSE} \h2 absorption data.  In Table~\ref{tab:Table1} we list  the target lines-of-sight (LOSs) selected from available {\it FUSE} and STIS UV absorption studies of \h2 \citep{Rachford2002,Sheffer2008} and \cuv $\lambda$2325 \citep{Sofia2004} towards more heavily reddened diffuse clouds. The target  distances in Table~\ref{tab:Table1}  are those available in   \cite{Sheffer2008} or in more recent papers \cite[e.g.][]{Jenkins2011}  and for a few are estimated from parallax given in the SIMBAD online data base. Though we do not use the distances  in our analysis we include them here to give an overall view of the distribution of our target selection.  The \hi column densities and their reference sources are also given in Table~\ref{tab:Table1}.  When available we use \hi  column densities derived from profile fitting of the naturally broadened wings of the Ly$\alpha$ absorption spectra \cite[e.g.][]{Diplas1994,Cartledge2004}.
%The extinction color excess E(B-V) values,  from Rachford et al. (2002) or Sheffer et al. (2008), for each target is also listed in Table 1.
 The \h2 rotational temperatures T$_{01}$(\h2) are from {\it FUSE} \h2 absorption data in \cite{Rachford2002,Rachford2009} and \cite{Sheffer2008}. For  HD 37021 \& HD 37061  we list  the typical  value $\sim$ 70 K. Although these two targets   are severely confused by the overall emission from the Orion molecular cloud, we include them here as examples of  PDR environments close to some of the {\it FUSE} target stars. We do not have the \h2 column density for these two target LOS but the \c+ column density  measurements are available from the  STIS UV absorption  \cite{Sofia2004, Sofia2011,Parvathi2012}. In Table 1 we list the \c+ column densities for 15 targets,   8 estimated from UV  \c+ absorption spectra (Sofia et al. 2004 \& 2011; Parvathi et al. 2012) and for 7 the \c+ column densities are derived from scaling O\,{\sc i}  and/or  S\,{\sc ii} \citep{Jenkins2011}.

The observations   of the fine-structure transition of C$^+$  ($^2$P$_{3/2}$ -- $^2$P$_{1/2}$) at 1900.5369 GHz, reported here  were made with  the HIFI band 7b   instrument \citep[][]{deGraauw2010} on {\it Herschel}  \citep[][]{Pilbratt2010}. The \cii spectra were obtained using the Wide Band Spectrometer with 0.22 \kmss velocity resolution  over 350 \kms.  For each target   we used the Load CHOP (HPOINT) with a sky reference at   $\sim$1\fdg4 to 2\degs off from the target.  To minimize the off source emission in the HPOINT spectra  of 4 targets  we used the OFF position   from the GOT C+ survey \cite[cf.][]{Pineda2013}.  In the case of  8 targets which had a second target close to it within 2\degs we used a common off position for each pair.  This choice helped us to constrain any off source emission which, if present, would appear equally in both HPOINT spectra. For 12 targets we observed two HPOINT spectra using an off position  at $\pm$1\degs in right ascension and  $\pm$1\degs in declination.

HIFI Band 7 utilized Hot Electron Bolometer (HEB) mixers which  produced strong electrical standing waves with characteristic periods of $\sim$320 MHz that depend on the signal power.  Prior to the release of HIPE-13 the  Level 2 \cii spectra  provided by the {\it Herschel} Science Center show these residual waves. Our data were processed in HIPE-14 using the standard  HIFI pipeline, which minimizes these residual waves.  We used a new feature in HIPE-14 that  mitigates the HEB electronic  standing wave in band 7 (at the \cii frequency) as part of the standard pipe line.  Any residual optical standing waves in the Level 2 spectra were removed  by applying the {\it fitHifiFringe} task to the Level 2 data to produce  satisfactory baselines.    The H-- and V--polarization data were processed separately and combined only after applying {\it fitHifiFringe}.   We also extracted the OFF-source spectrum to examine and correct for any \cii spectral contamination due to the OFF--source subtraction.

For three targets (HD 115071, HD 168076 \& HD 170740) which are close to the Galactic plane ($|b|$ $<$ 1\fdg0)  we observed an OTF  6 arcmin longitudinal scan centered at the target star positions.  All HIFI OTF scans were made in the LOAD-CHOP mode using a reference off--source position from the GOT C+ survey \cite[cf.][]{Langer2014_II} which were about 1 to 2 degrees  away in latitude.    The OTF  6 arcmin scans are sampled every 20 arcsec and the total duration of each OTF scan was typically $\sim$2000 sec.

%%% Figure 1
 % \newpage
\begin{figure*}[!ht]
\hspace{-1.5cm}
%\centering
\includegraphics[scale=0.85]{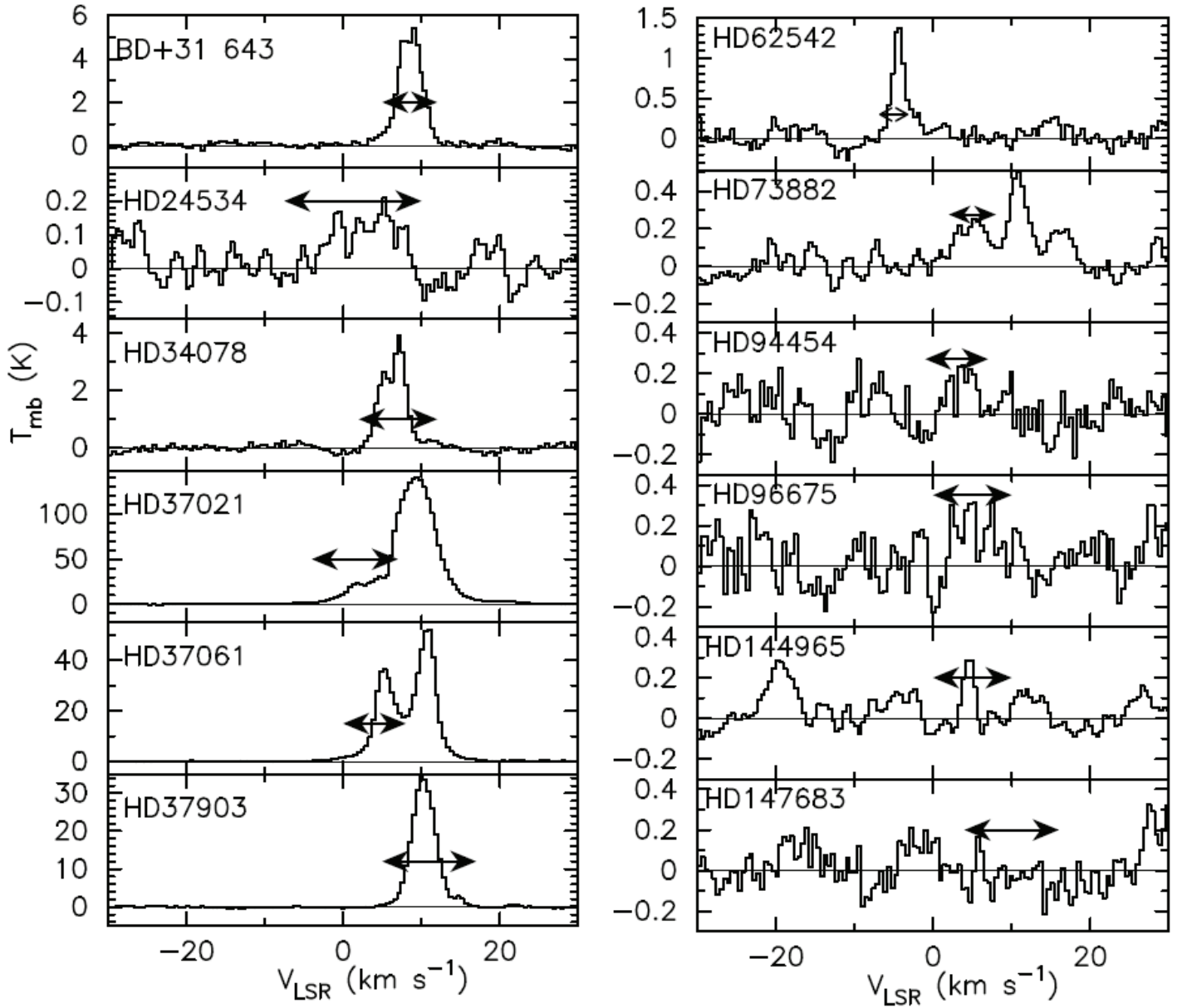}
\caption{HIFI pointed observation spectra of \cii 158 \microns emission towards  12 {\it FUSE} target stars with names in the upper left corner of each spectrum.    The horizontal  double arrows mark the V$_{LSR}$ range of  CH, CH$^+$ or atomic line absorption velocities available in an online catalog (Welty 2016) except for  HD 73882    \citep{Ritchey2011,Lacour2005}.
  }
\label{fig:f1}
\end{figure*}

%Figure 2
%\newpage
\begin{figure*}[!ht]
\hspace{-1.5cm}
%\centering
\includegraphics[scale=0.85]{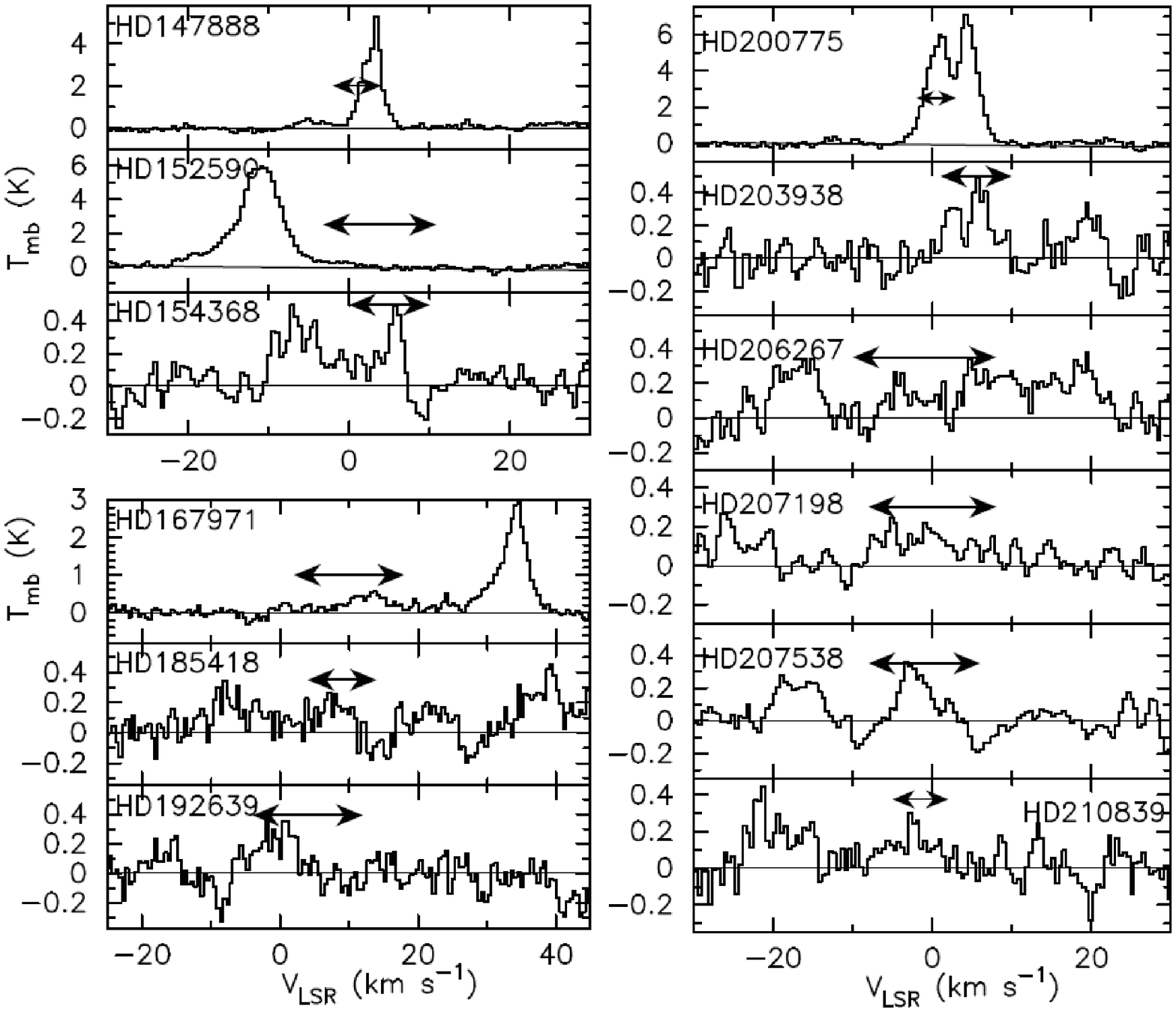}
\caption{HIFI pointed observation spectra of \cii 158 \microns emission towards  12 {\it FUSE} target stars with names in the upper left corner of each spectrum.   The horizontal   double arrows mark the  V$_{LSR}$ range of  CH, CH$^+$ or atomic line absorption velocities available in an online catalog (Welty 2016)  or in other resources, such as: \cite{Ritchey2011} for HD 154368; \cite{Lacour2005} for HD 185418, HD 192639, HD 206267 \& HD 207538; \cite{Federman1997} for HD200775. }
\label{fig:f2}
\end{figure*}
  We processed the OTF scan map data following the procedure discussed in \citet[][]{Velusamy2014}. Using the HIPE-14 Level 2 data (which are already corrected for the HEB standing waves) the \cii maps were made as ``spectral line cubes'' using the standard mapping scripts in HIPE. Any residual  HEB and optical standing waves in the ``gridded'' map spectra were minimized further by  applying  {\it fitHifiFringe}   to the individual   spectra in the gridded scan map.  %We took the additional precaution in {\it fitHifiFringe} of disabling  {\it DoAverage}  in order not to bias the spectral line ``window''.
   The H-- and V--polarization data were processed separately and were combined only after applying {\it fitHifiFringe} to the gridded data. This approach minimizes the standing wave residues in the scan maps  by taking into account   the standing wave  differences between H-- and V--polarization.     We then  used the processed spectral line data cubes to make longitude--velocity ($l$--$V$) maps of   the \cii emission as a function of the longitude range in  each of the 3  OTF scan observations.     For HIFI observations we used the Wide Band Spectrometer (WBS) with a spectral resolution of 1.1 MHz for all the scan maps. The final $l$--$V$ maps presented here were restored with a velocity resolution of 1 \kms.
%Figure 3
\begin{figure}[!ht]
%\centering
\hspace{-1.5cm}
\includegraphics[scale=0.525]{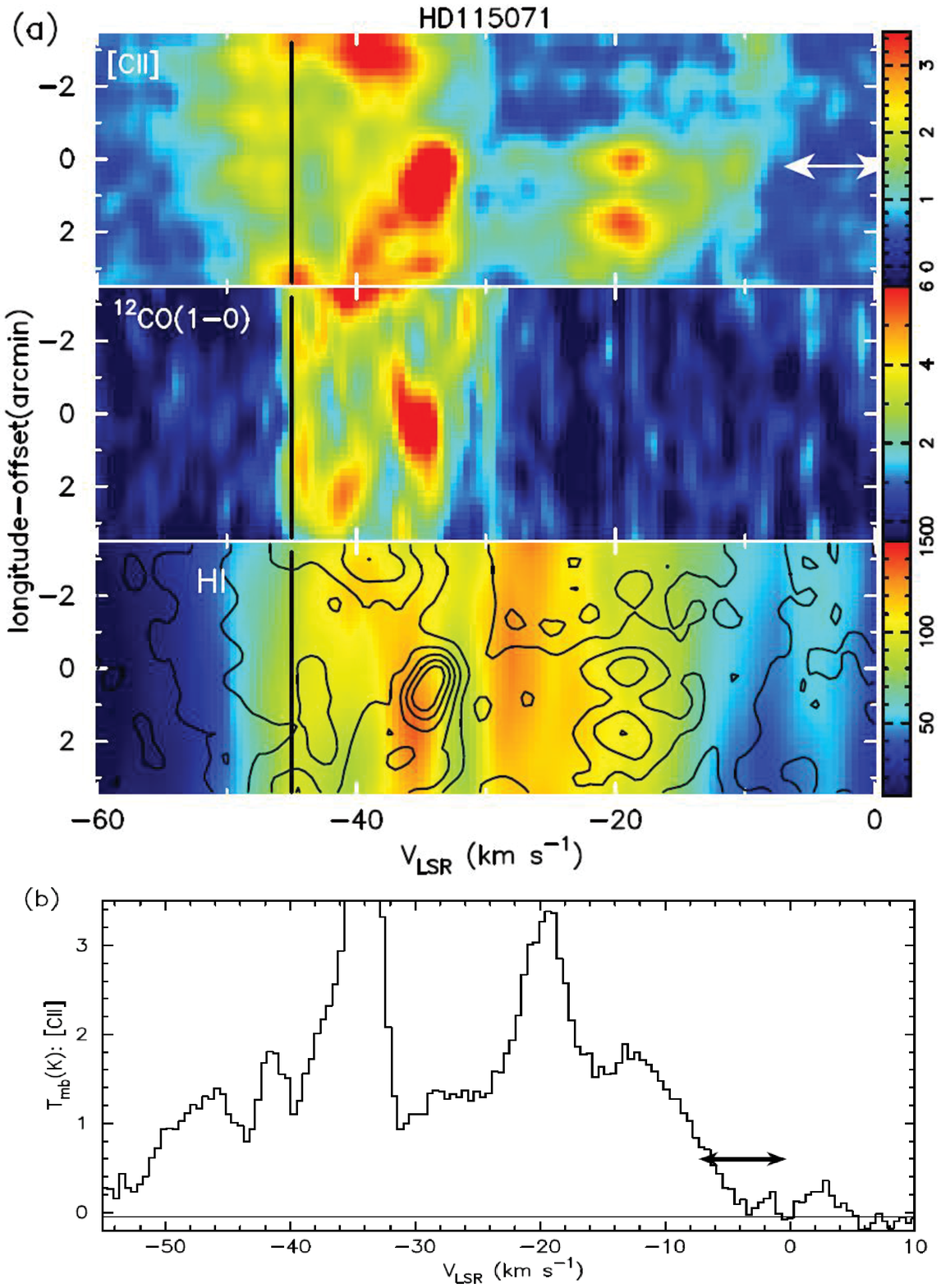}
\caption{HD 115071 (G305.764-0.152): {\it (a)} {\it l--V} maps of \cii 158 \microns emission   \his, and CO.   The  position offset zero corresponds to the target star. The  \hi 21 cm  data are from the SGPS  \citep[][]{McClure2005}, \co(1-0)  data are taken from the ThrUMMS survey \citep[][]{Barnes2011}.
Longitude-offset zero corresponds to $l$ = 305.7642\deg.  The black vertical line represents the tangent velocity (see text).    The brightness (T$_{mb}$ K)  scales for the images are indicated by the color bars.  \cii emission contours are overlaid on the image of the  \hi 21 cm emission and the contour levels (T$_{mb}$) are 1, 2, 3, 4, 5, and 6 K.  {\it b:} \cii  spectrum towards the target star (at zero longitude-offset) observed by HIFI in the pointed observation.
 %The arrow marks the target V$_{LSR}$  and the dashed line indicates the V$_{LSR}$ of the heliocentric zero velocity.
The horizontal double arrow marks the V$_{LSR}$ range of  CH absorption \cite[from][]{Andersson2002}. }
\label{fig:f3}
\end{figure}

 At 1.9 THz the angular resolution of the {\it Herschel} telescope is 12\arcsec, but  the \cii OTF observations  employed   20\arcsecs sampling. Such fast scanning results in an undersampled scan, and  broadening of the effective beam size along the scan direction  \citep[][]{mangum2007}.  Therefore all \cii maps have been    restored  with an effective beam size corresponding to  twice the 20\arcsecs sampling interval along the scan direction. Thus the shorter integration time per pixel  in the final OTF maps {restored} with an 40\arcsecs beam  and 1 \kmss wide channels yields an rms $\sim$0.22 K for T$_{mb}$  which is slightly larger than in the HPOINT mode.

 To compare the distribution of the C$^+$ gas with the atomic and molecular gas we use the \co and   \hi 21 cm data  in the southern Galactic plane surveys available in  public archives.  For HD 115071  the \cos(1-0)  data are taken from the Three-mm Ultimate Mopra Milky Way Survey\footnote{www.astro.ufl.edu/thrumms. The data are from the Mopra radio telescope, a part of the Australia Telescope National Facility which is funded by the Commonwealth of Australia for operation as a National Facility managed by CSIRO. The University of New South Wales (UNSW) digital filter bank (the UNSW-MOPS) used for the observations with Mopra was provided with support from the Australian Research Council (ARC), UNSW, Sydney and Monash Universities, as well as the CSIRO} (ThrUMMS) observed with the 22m Mopra telescope  \citep[][]{Barnes2011} and the   \hi 21 cm data are taken from the Southern Galactic Plane Survey (SGPS) observed with the Australia Telescope Compact Array  \citep[][]{McClure2005}.  For the other two targets we use   \hi 21 cm data from the  VLA Galactic plane survey (VGPS)  \citep[][]{Stil2006} and the $^{13}$CO(1-0) data from the Galactic Ring Survey (GRS) \citep{Jackson2006}.

\section{Results}
\label{sec:results}
\subsection{HIFI \cii emission spectra and {\it (l--V)} maps}
\begin{figure}[!t]
%\centering
\hspace{-0.2cm}
\includegraphics[scale=0.6, angle =-90]{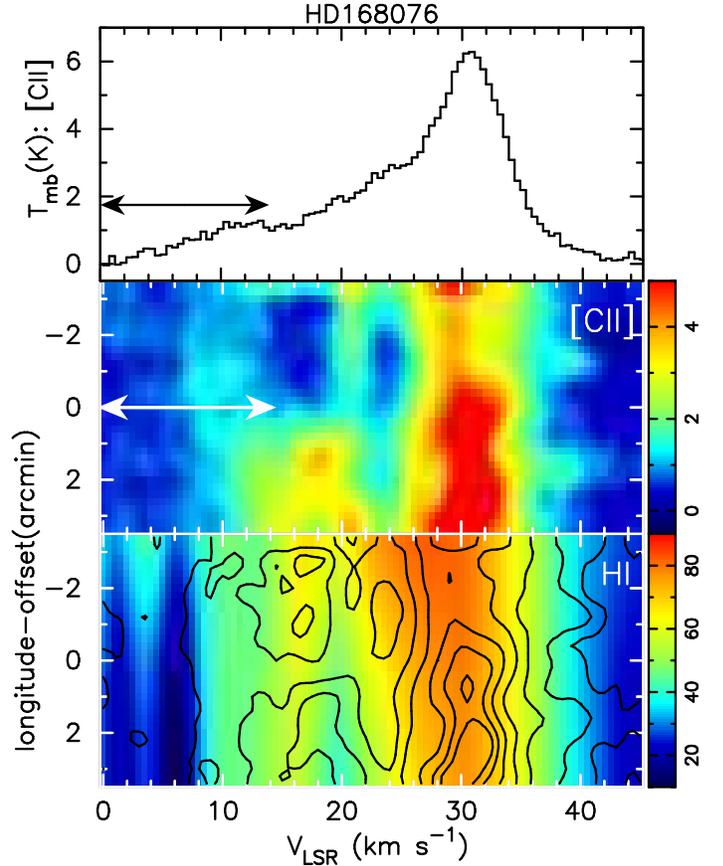}
\caption{HD 168076 (G016.937+0.837):  {\it Top panel:} HIFI  \cii velocity spectrum towards the target star (at zero longitude--offset)  in a pointed observation.  The double arrow marks the $V_{LSR}$ range of the \ki absorption (Welty 2016 online catalog). {\it Image panels:} {\it l--V} maps of \cii 158 \microns emission  along with   \hi 21 cm. No CO map data are available for this target LOS.  Longitude-offset zero corresponds to $l$ = 16.9373\deg.  The    \hi 21 cm data are from the VGPS  \citep[][]{Stil2006}   observed with a beam of 1\arcmin.    The brightness (T$_{mb}$ K)  scales for the images are indicated by the color bars.}
\label{fig:hd168076}
%\vspace{-0.5cm}
\end{figure}

\begin{figure}[]
%5\centering
\hspace{-1.5cm}
\includegraphics[width=12cm]{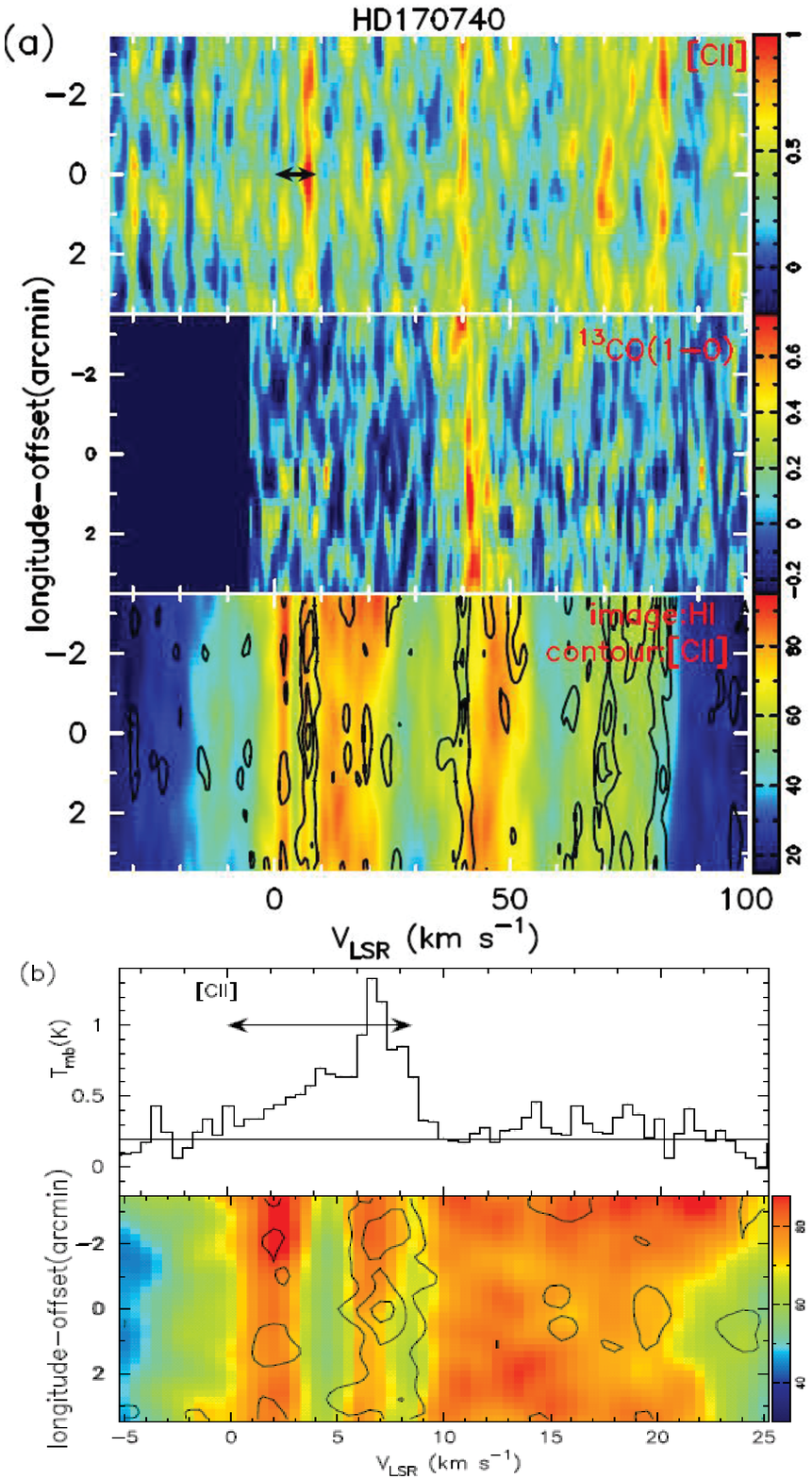}
\caption{HD 170740 (G021.057-0.526):  (a)  {\it l--V} maps of \cii 158 \microns emission (top) along with $^{13}$CO (middle) and   \hi 21 cm (lower).    The  \hi  data are from the VGPS  \citep[][]{Stil2006} and \13co(1-0) from the GRS \citep{Jackson2006}. Longitude-offset zero corresponds to $l$ = 21.0574\deg.    The brightness (T$_{mb}$ K)  scales for the images are indicated by the color bars.  (b)  {\it Top panel:} The HIFI  \cii velocity spectrum towards the target star (at zero longitude-offset) observed in a pointed observation.  The double arrow marks the $V_{LSR}$ range of the CH absorption (Ritchey et al. 2011). {\it Image panel:} Enlarged view of the {\it l--V} maps for the selected velocity range. The \cii intensity contours are overlaid  on  the   \hi 21 cm  map.}
\label{fig:hd170740}
\end{figure}

 The HIFI spectra of the velocity resolved \cii far-IR spectral line emission towards all targets in our sample are shown in Figures~\ref{fig:f1} to \ref{fig:hd170740}.  The spectra are shown as main beam temperature, T$_{mb}$, plotted against the local standard of rest velocity, $V_{LSR}$, (we use a Band 7 main beam efficiency $\eta_{mb}$ = 0.585 and forward efficiency $\eta_l$ = 0.96)\footnote{HIFI Release Note HIFI-ICC-RP-2014-001}.  In these spectra, except for a few target LOSs, we detect weak to strong \cii emission features at multiple velocities.  Unfortunately, the {\it FUSE} UV \h2 absorption data do not have velocity resolved spectral information to make direct comparison of the velocity features in the \cii emission spectra with the \h2 gas seen in UV absorption.    However, to identify $V_{LSR}$ velocities of the \cii emission from the \h2 gas seen in absorption we can use available velocity resolved atomic and molecular UV absorption spectra towards our target stars, because these species are associated with diffuse gas.  Here we primarily use CH or CH$^+$, and if these are not available, atomic \oi, Na\,{\sc i}, or K\,{\sc i} absorption spectra.  We can then associate the $V_{LSR}$  velocity ranges over which absorption is detected in any of these spectra with the corresponding HIFI \cii emission spectral features.  In the spectra in Figures~\ref{fig:f1} to \ref{fig:hd170740} the UV absorption velocity range for each target derived from these tracers is marked by a horizontal double arrow.  For most target stars we use the velocity resolved absorption data available in the online catalog of spectra\footnote{http://astro.uchicago.edu/$\sim$dwelty/ew-atom.html}  compiled by \cite{Welty2016} and the references therein. Whenever possible we use the high-resolution CH $\lambda$4300 line absorption spectra because, like most hydrides, it is present in diffuse molecular clouds \cite[cf.][]{Gerin2016} and shows a linear relationship between CH and \h2 column densities for A$_V$ $<$ 4 mag. \cite[e.g.][]{Federman1982,Rachford2002,Sheffer2008}.  Furthermore, for the \cii spectra in a few targets  the identified V$_{LSR}$ velocities  near V$_{LSR}$=0 extend to  positive and negative values which appear inconsistent with predicted velocity ranges for kinematic distances towards these target LOS if we assume circular Galactic orbits and a simple Galactic rotation curve.  However, if we adopt the V$_{LSR}$ velocities inferred from available directly observed UV absorption spectra it eliminates the need to model non-circular or peculiar velocities which will likely introduce more uncertainties.

 By selecting only the absorption velocity range of the tracers of \h2 gas we can thus exclude any \cii emission features along the LOS that are not associated with the \h2 gas observed by {\it FUSE}.   However, this exclusion is true only for  LOS toward the outer Galaxy because we obtain a unique kinematic distance for any given V$_{LSR}$ using the V$_{LSR}$--distance relationship derived by assuming circular Galactic orbits and using a simple Galactic rotation curve. For the  LOS toward the inner Galaxy there is degeneracy in the V$_{LSR}$ -distance relationship resulting in a near- and far-distance solution \cite[e.g.][]{Duval2009}.  In our {\it FUSE}--\h2 sample, of the 26 Galactic LOSs with \cii detections, 14 LOSs are toward the outer Galaxy with no distance ambiguity in associating \cii emission velocity with \h2 gas, while  12 are towards the inner Galaxy which are affected by the near- and far-distance ambiguity in  the location of \cii emission whether it is   between the target star and observer.     Thus  the V$_{LSR}$ velocities, marked in Figs. 1 to 5,  provide  distances to  two regions along each LOS, one at the far-distance (typically $>$ 10 kpc)  beyond the target star   and the second region at  near-distance between the star and observer. In these target sights the \cii] emission detected at a given $V_{LSR}$  can originate from the gas at the near-distance (between the target and observer) or at the far-distance beyond the target  or may include emission from both near- and far-distance sources.   We calculate  the near- and far-distances for the $V_{LSR}$  velocities shown in the \cii spectra in Figs 1 to 5 and estimate their $z$-distance above or below the Galactic plane using their LOS Galactic latitude and far-distances which are typically $>$ 10 kpc. Except for HD 115071 and HD 192639, in all other target LOSs any gas at  the far-distance would be located at $|z|$ $>$ 150 pc  (up to a few kpc in many cases)   where little diffuse \h2 gas is likely be present  (\cite{Velusamy2014} derived a scale height of 100 pc for diffuse \h2 using the GOT C+ \cii survey--see their Figure 15 and Table 3). Therefore, we conclude that for all but two LOS, the \cii emission identified by the $V_{LSR}$  velocities do not contain much, if any, contamination from any other sources beyond the target stars.  In the case of HD 115071 and HD 192639 the far-distance gas if present would be at $z$ = 25 pc and 100 pc, respectively.   If the emissions   in the near- and far distances  are roughly equal,  the \cii intensities, which are assumed to be in the near-distance in our analysis,  are overestimated by a factor of two.  In this case the thermal pressures derived from \cii intensities for these two LOSs (Table 2) assuming all emission is in front of the target star are higher than the mean value for thermal pressure (see Section 4.1). On the other hand,  if we have to correct their \cii intensities for contribution from far-distance gas their thermal pressures would be lower but still  consistent with our results and conclusions.

In each spectrum in Figs. 1 to 5 the baseline used is marked by the horizontal line  through zero intensity. These baselines were first fitted in  the HIPE data processing using the full useable velocity range ($>$ 160 \kms) as part of the {\it fitHifiFringe} task which also removes the residual standing waves in the spectra. For a few LOSs the baselines were then improved by fitting over a narrower velocity range ($\sim$ 50 \kms) centered near the V$_{LSR}$ of the \h2 gas.  We caution that  some of the low level  velocity features (T$_{mb}$ $<$ 0.2 K) in the spectra  may be residual standing waves and may not be real.   It may be noted that the rms noise estimates (in Table 2) exclude  the low level standing waves.

For three  {\it FUSE} targets: HD 115071 (G305.764+0.152), HD 168076 (G016.937+0.837) and HD 170740 (G021.057-0.526),  for which we have \cii HIFI OTF scan maps, we show   pointed spectra and  {\it (l--V)} maps in Figures~\ref{fig:f3} to \ref{fig:hd170740}.  These three targets have low Galactic latitudes (within 1\degs of the plane) and therefore for these regions there exist high angular resolution ancillary   \hi 21 cm  and some CO map data. The \cii maps allow  us to resolve spatially the diffuse \h2 gas detected in the LOS UV absorption spectra.
The availability of the    \hi 21 cm  and CO maps allow us to further characterize the diffuse \h2 molecular cloud.  The   {\it (l--V)} maps of \cii   intensity in T$_{mb}$  and the corresponding  \hi 21 cm  and CO maps (when available) are shown  as color maps in $T_{mb}$.  %on a longitude--offset from the target versus V$_{LSR}$.  The color wedges shown to the right of the images indicate intensity scale.
The spectra shown in Figures~\ref{fig:f3} to \ref{fig:hd170740} have lower rms noise than those in the {\it (l--V)} maps because of the longer integration used for the pointed observations compared to the OTF scans. Furthermore, \cii emission in these pointed spectra correspond to a narrow beam (12\arcsec$\times$12\arcsec) while the {\it (l--V)} maps have a larger beam  size of (40\arcsec$\times$12\arcsecs) (see Section~\ref{sec:observations}).

All three {\it (l--V)} maps show a clear detection of \cii at the  velocities of the \h2 absorption.  In addition, we  detect  \cii features at velocities corresponding to the inner Galaxy and many have associated CO counterparts. However, for both HD 115071 and HD 170740 there is no detectable CO matching the \h2 absorption clouds, confirming that these are diffuse clouds with very little  or no CO in them.  Here we discuss only the \cii features associated with the {\it FUSE} \h2 absorption.
%Figure 4

%Figure 5
\subsection{\cii intensity analysis and  \h2 cloud properties}
A direct method to measure the density and temperature in \h2 clouds, and therefore their thermal pressure, is by observing several transitions of one or more gas probes that have the right energy spacing and collisional rate coefficients that probe the relevant parameter range.
\c+ has a fine-structure transition at 158 \microns that arises from the $^2$P$_{3/2}$ level at an energy of 91.21 K, comparable to the kinetic temperatures in the diffuse and translucent clouds.  It has been used to study the physical conditions in the ISM across the Galaxy \cite[cf.][]{Pineda2013,Langer2014_II} as well as in the Galactic Center \cite[cf.][]{Garcia2016, Langer2017}.
Because \c+ has only one fine-structure transition, \cii emission  can only be used to derive one of   three key physical parameters of the gas, either density, kinetic temperature, or column density as a function of the other two.
For example, Gerin et al. (2015) derive the thermal pressures using C$^+$ column densities measured from the \cii absorption data, with auxiliary assumptions about density, temperature, and the \hi source.
 However, in the case of   \cii observed in emission, as shown by \cite{Langer2014_II}, there are regimes in density--temperature space where \cii can be used to estimate the product of density and temperature, or the thermal pressure, $P_{th}$ = $nT$, if one knows the \c+ or \h2 column density. (We adopt  the notation $P_{th}$  in units of K cm$^{-3}$ to express pressure divided by Boltzmann's constant).   In an Appendix  we extend the treatment of \cite{Langer2014_II} and give a general solution applicable to   a wider range of cloud conditions.
  Here we employ this approach as it applies to the lines of sight to the  {\it FUSE} target stars and compare the pressures derived from \cii emission to those using UV absorption lines.

\subsubsection{  \cii intensities in the \hi  and \h2 gas components }
To determine the \cii spectral line intensities it is critical to use a $V_{LSR}$ window that realistically represents the range of \h2 absorption within the \cii velocity profile. The ranges over which the velocity integrated \cii intensities were obtained are indicated in the spectral line velocity profiles in Figures~\ref{fig:f1} to \ref{fig:hd170740}.  In cases where the \cii velocity feature is blended with adjacent emissions (e.g. HD37021, HD37061HD 152590, HD 200775) we use multi-Gaussian fits to compute the integrated intensities. The integrated \cii intensities, $I$(\ciis) in K km s$^{-1}$, are given in column 3 in Table 2.   $I$(\ciis) is detected at the  5-$\sigma$ level  or better   for most targets,  for three at the 3- to 4-$\sigma$ level, with one non-detection for HD 147683.     In a few cases the \cii detections are marginal considering the presence of other comparable velocity features in the spectra. Nevertheless, they are statistically significant for studying the thermal pressures as the marginal detections can be used to determine upper limits to thermal pressure.

 The \cii intensities integrated over the $V_{LSR}$  ranges of the \h2 UV absorption, in addition to the emission from the \h2 cloud (\c+ excited by \h2 collisions), includes some emission from the \hi  gas along the line of sight (\c+ excited by H collisions). We can thus write the observed \cii intensity as
\begin{equation}
I([CII])_{obs}  = \int T_{mb} dV =  I([CII])_{\rm HI}  +  I([CII])_{\rm H_2}
\end{equation}
where $I$(\ciis)$_{HI}$  and $I$(\ciis)$_{H_2}$ are the emission in the \hi  and \h2 clouds respectively.
To analyze the \cii emission from just the \h2 cloud  we follow the steps below: (1) Determine $I$(\ciis)$_{obs}$ by integrating \cii emission over the $V_{LSR}$ window; (2) Calculate  $I$(\ciis)$_{HI}$ from $N$(\his) with an assumed \c+ fractional abundance and cloud pressure; (3) Calculate $I$(\ciis)$_{H_2}$   correcting the observed \cii intensities for any contribution from the \hi  gas; and finally, (4) Calculate the thermal pressure $P_{th}$ and density n(\h2) using the \cii intensity in the \h2 gas, the measured \h2 column density and temperature.

We use the observed \hi  column densities listed in Table 1 to determine the \cii contribution from the \hi clouds. In Figure 6 we show a scatter plot of the total \cii intensity ($I$(\ciis)$_{obs}$)  against the \hi  column density $N$(\his). We do not see any  strong correlation to suggest that there is  any dominant contribution to $I({\rm CII})_{obs}$ from the \hi  column densities. However, the lower envelope of the scatter plot  suggests there is a small contribution to $I({\rm CII})_{obs}$ that increases slowly with $N$(\his) as indicated by the solid line in Figure 6.  We can therefore use Eq. A.9 to quantify the \hi  contribution in terms of \hi  gas pressure  and  $N$(\his) as a proxy for $N$(\c+) in the \hi  clouds. Here we adopt a value for \c+ fractional abundance $x$(\c+/\his) = $x$(C/H) = 2.2 $\times$10$^{-4}$ (see section 4.2).

% Figure 6
\begin{figure}[!t]
%\hspace{-0.5cm}
\centering
\includegraphics[scale=0.35, angle = -90]{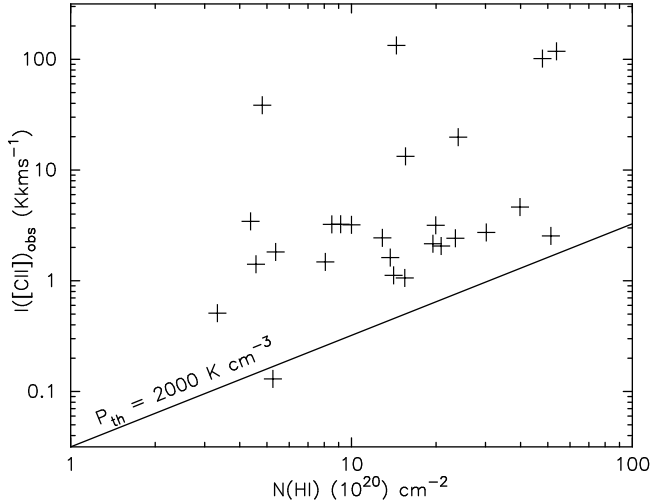}
\hspace{-0.5cm}
\caption{ Scatter plot of the observed HIFI \cii intensity  against  the \hi  column density. The solid line shows a lower boundary to the \cii intensity as a function $N$(\his).  The solid line also represents   \cii intensities for an assumed  thermal pressure of the lower boundary   $P_{th}$ = 2000 K cm$^{-3}$  (in Eq. A.9  assuming  x(C$^+$/\his) = 2.2$\times$10$^{-4}$  -see text).  }
\label{fig:f6}
\end{figure}

 \cite{Jenkins2001} derived the thermal pressure in diffuse atomic clouds from absorption line measurements of neutral carbon towards 26 stars and find an average pressure of 2240 K cm$^{-3}$.  In a recent survey towards 89 stars \cite{Jenkins2011} (hereafter referred to as JT2011) derive their thermal pressure using the UV \ci absorption.  The number distribution of their pressure estimates have a mean value of $P_{th}$ = 3800 K cm$^{-3}$.  Because \ci is present in \h2 molecular gas rather than in the atomic \hi  gas this value is more representative of diffuse molecular clouds.  However, 10\% of their sample (about 9 target LOSs) have much lower values in the range of 900 to 2500 with a mean of value of 2000 K cm$^{-3}$. Since the pressure in the \hi  gas is likely to be lower  we assume the lower values in the JT2011 pressure distribution and for our analysis of \cii in \hi  gas adopt $P_{th}$ $\sim$2000 K cm$^{-3}$. The $I({\rm CII})_{\rm HI}$ computed using this value for thermal pressure in \hi  as a function of  $N$(\his) is shown in Figure 6 as the solid line which also fits well the lower envelope of the   observed \cii intensity versus $N$(\his). To determine $I$(\ciis)$_{HI}$  we use the low density solution (Eq. A.9). Thus,
 using a gas pressure $P_{th}$ = 2000 K cm$^{-3}$ for atomic clouds and expressing \hi column density in units of 10$^{20}$ cm$^{-2}$,
  we can reduce Eq. A.9 to,
\begin{equation}
I({\rm CII})_{\rm HI} = 3.3\times10^{-2} N_{20}({\rm HI})\qquad\qquad{\rm K\, km\, s^{-1}}.
\end{equation}

  For each target LOS we estimate the ${\rm I}({\rm CII})_{HI}$   in the \hi  gas component from Eq. 2 using the respective \hi  column densities  in Table 1 column 4.   In principle we can now compute the $I({\rm CII})_{\rm H_2}$ using these values for  $I({\rm CII})_{HI}$.  However, it should be noted that the \hi  column density used in Eq. 2 corresponds to the entire path length between observer and the target star while the observed velocity integrated \cii emission is over a smaller path length limited  to the narrow  V$_{LSR}$ range of the UV absorption features.  Ideally we should use the \hi  column density integrated over the same  V$_{LSR}$ range used to measure the \cii intensity.  The \hi  column density derived from extinction or $L\alpha$ observations represents an integration over a much wider V$_{LSR}$ range, corresponding to that between heliocentric zero to target radial velocity while the diffuse \h2 cloud is located over a narrower velocity range.  Therefore the \hi  contribution within the V$_{LSR}$ range of the \cii emission is overestimated.  Indeed for the three LOS for which we have high angular resolution \hi  spectra (Figs. 3 to 5) the fraction of $N$(\his) seen in  extinction   within the \cii emission  V$_{LSR}$  range is 68\% (HD 170740), 12\% (HD168076), and 60\% (HD115071) of the total velocity range to the target.  The \cii intensity from the \hi  for all our LOSs varies between 0.1 K \kmss to 1.7 K \kms.  However it is not likely that all of this \hi  contribution is overlapping the \h2 gas component seen in the \cii emission. We can then assume the true value of the  \cii intensity in the \h2 gas is between the total observed  and that corrected for \hi  contribution; here we use the average value:
\begin{equation}
I({\rm CII})_{H_2} = I({\rm CII})_{obs} - 0.5\times I({\rm CII})_{HI}
\end{equation}

  For the three LOSs for which we have higher angular resolution \hi spectra the full (100\%) ${\rm I}({\rm CII})_{HI}$ is subtracted. In Table 2 we give the observed \cii intensities integrated over the  V$_{LSR}$  range as indicated in the spectra in Figs. 1 to 5, \cii intensity from the \hi gas component and the \hi  corrected \cii intensities  (Eq. 3) in the \h2 gas, ${\rm I}({\rm CII})_{H_2} $. Half of the contribution  in the \hi  gas is added to the uncertainty in the estimate of I(CII)$_{H_2}$.  The 1-$\sigma$ uncertainty in \cii intensity from rms noise  alone  is in the range 0.2 to 0.5 K \kms. The 1-$\sigma$ error in Table 2 includes the uncertainties from the  noise and the estimated \hi  contribution.

\subsubsection{  \cii cloud intensities and \h2 cloud pressure \& density}

In Figure 7 we show   the measured $I$(\ciis)$_{H_2}$  for each target LOS plotted against the respective $N$(\h2).  The data in  Fig. 7 show some scatter in the observed \cii emission  for any given \h2 column density in the diffuse cloud. The lack of a tight correlation between the \cii intensities and \h2 column densities may be the result of a wide range of pressures in the individual LOSs and  at high \cii intensities due to the influence of the target star itself creating a high pressure PDR environment.     To understand better the scatter in Fig. 7 we estimate the predicted $I$(\ciis)$_{H_2}$ as a function of $N$(\h2) using Eq. A.8 (for the low density solution in Appendix A  and using $N$(\h2) as proxy for $N$(\c+)) for several values of thermal pressure in \h2 gas.   As discussed in Section 4.1   in   our $I$(\ciis)$_{H_2}$  -- $N$(\h2)  analysis, we use a constant value of $x$(C/\h2)= $4.4\times 10^{-4}$   for all target LOS to estimate $N$(\c+) from the measured $N$(\h2). The solid lines in Figure 7 delineate the lines of constant thermal pressure in  $I$(\ciis)$_{H_2}$ -- $N$(\h2) space.  For a majority of the LOSs the \cii intensities are consistent with those expected for thermal pressures between 5000 and 20000 K cm$^{-3}$ with a few high pressure \h2 clouds possibly located in a PDR environment excited by the target stars.

To analyze the thermal pressures and their distribution in our {\it FUSE} \h2 sample we determine the  values of $P_{th}$ in each LOS. Since we have estimates of the \h2 gas temperatures  (T$_{01}$) from the excited levels of \h2, $T_{01}$(\h2) for this analysis we do not use the the low density solution in Appendix A. Instead we  use  the general  solution  Eq. B.2 (from Appendix B) for the thermal gas pressure and density, substituting $N$(\c+) = $4.4\times 10^{-4}N$(\h2), temperature functions $f$(T), $b$(T) with corresponding values for T$_{01}$ and using  n$_{cr}$(\h2)= 4500 cm$^{-3}$, and $a_0$ =3.28$\times$10$^{-16}$:
\begin{equation}
P_{th} =   \frac{150\times{\rm I}({\rm CII})_{H_2}}{f(T)N_{20}(H_2)}\frac{1}{1- 0.07\times b(T)I([{\rm CII}])_{H_2}/ N_{20}({\rm H_2})}
\end{equation}
\noindent where $P_{th}$ is the cloud pressure expressed  in K cm$^{-3}$ and the temperature functions are
$f(T) = {T_{01}}^{-1}e^{-91.21/T_{01}}$; and
$b(T) = 1+0.5e^{91.21/T_{01}}$.
Finally,  the volume density  $n$(\h2) is derived from the $P_{th}$ obtained from the general solution as
\begin{equation}
n(H_2) = P_{th}/T_{01}  \qquad\quad {\rm cm^{-3}}.
\end{equation}
The derived pressures and densities are summarized in Table 2.  The uncertainty includes only that in the \cii intensity and does not include the uncertainty in the \h2 column density, kinetic temperature, or \c+ fractional abundance. In most cases the overall uncertainty is $\sim$ 20\%.   Note that the first term in Eq 4 is the same as the solution for the low density approximation.   For 3 out of the total 26 LOSs the \cii intensities are too high, $I$(\ciis)$_{H_2}$/$N_{20}$(\h2) is $>$ 7, and for values beyond  7 the general solution (Eq. 4) fails. Therefore, for these three target LOSs (HD 37021, HD 37061 \& HD 37903) we show the values derived from the low density approximation which gives a lower bound. These and a few other     high pressure LOSs are identified with the spectral type of  target star which is also the exciting source of a PDR environment as discussed in Section 4.3.    In general as shown in Appendix B,  the   thermal pressure  derived from  the \cii emission intensity analysis used in this paper is also consistent with radiative transfer analysis.
\begin{figure}[!t]
%\hspace{-0.75cm}
\centering
\includegraphics[scale=0.35, angle = -90]{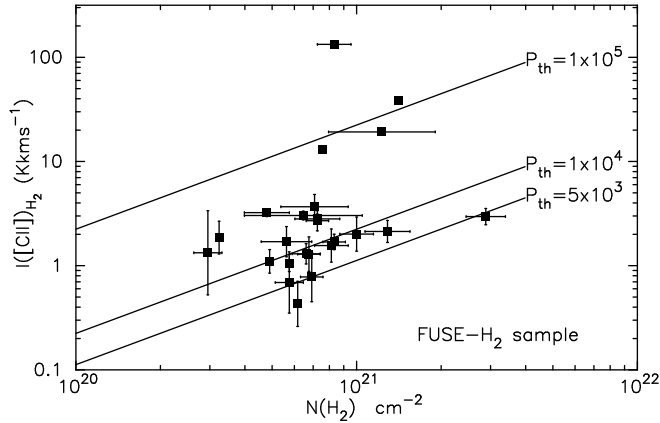}
%\hspace{-0.5cm}
\caption{ 158 \microns  \cii intensity from the \h2 gas as a function of  \h2 column density.     The predicted \cii intensity for a given \h2 column density for a range of molecular  gas pressure is also plotted. The error bars  for the \cii intensities include the uncertainty in subtracting the \hi  contribution and the noise in the spectrum.
The solid lines show the expected \cii intensity as a function of \h2 column density for assumed thermal pressures (K \kms;  Eq. A.8 in Appendix A) adopting  x(\c+/H$_2$) = 4.4$\times$10$^{-4}$.
% TBD plot error in N(\h2)
}
\label{fig:f7}
\end{figure}
\begin{figure*}[!t]
%\hspace{-0.5cm}
\centering
\includegraphics[scale=0.575, angle = -90]{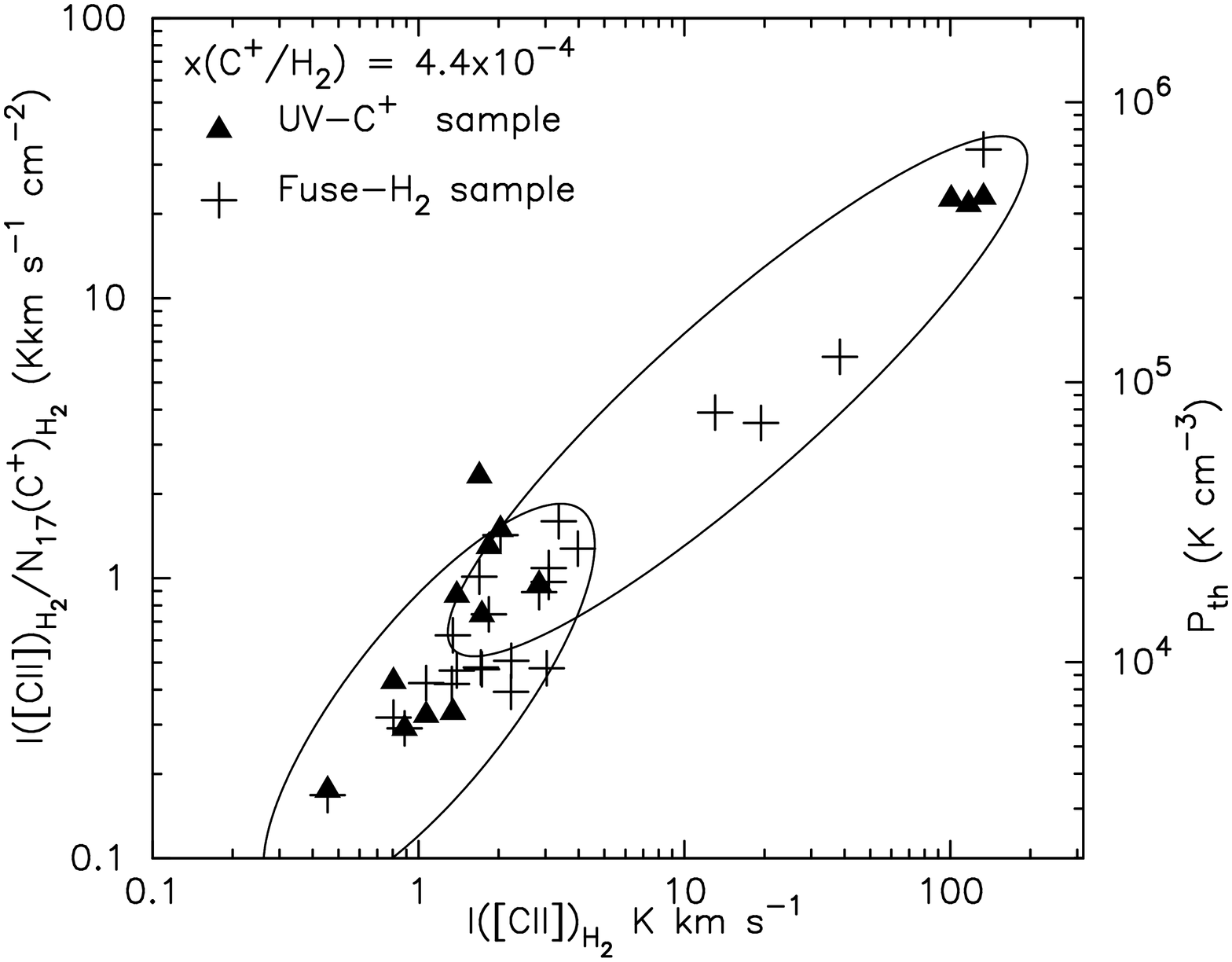}
%\vspace{-1cm}
\caption{  $I$(\ciis)$_{H_2}$/$N$(\c+)$_{H_2}$   versus  $I$(\ciis)$_{H_2}$ using the data in Tables 1 \& 2 (see text).  $I$(\ciis)$_{H_2}$/$N$(\c+)$_{H_2}$ is a direct measure of thermal pressure, $P_{th}$, and the corresponding values (see Eq. 8) are marked on the Y-axis to the right.  The crosses    represent data points ({\it FUSE}--\h2 sample) for which the $N$(\c+)$_{H_2}$ is  computed from  $N$(\h2) using x(\c+/\h2)= 4.4$\times$10$^{-4}$.  Note that for the crosses we use the \h2 column density  and the corresponding \cii intensity in the \h2 gas component along the LOS.   The  filled triangles   represent data points (UV--\c+ sample) for which the $N$(\c+)$_{H_2}$  is obtained using $N$(\c+) from direct UV absorption studies (references in Table 1) and \h2 fraction $f$(\h2) fraction (see text).  The  ellipses show the distribution of the \h2 clouds in the  $I$(\ciis) versus  $P_{th}$ space from the lower left to upper right suggesting a population of clouds in an evolutionary sequence-- from low density, low pressure to a high pressure regime.
%\vspace{0.5cm}
 }
\label{fig:f8}
\end{figure*}
%\vspace{0.5cm}

 \begin{table*}[!th]
 %\renewcommand{\arraystretch}{0.8}
%\centering
\begin{center}
%\hspace{-1cm}
\caption{HIFI \cii intensities and derived thermal pressures and densities: Comparison with UV \ci and CO results}
\renewcommand{\tabcolsep}{0.25cm}
\begin{tabular}  {l  c c c c c c c l}
\hline
\hline
Target	&	 	$I$(\ciis)		&	$I$(\ciis)	&	$I$(\ciis)			&	$P_{th}$$^1$		& $P_{th}$$^2$ & n(\h2)$^1$	&  n(\h2)$^3$	& Notes	\\
Star	&	 	K km s$^{-1}$			&	K km s$^{-1}$	&	K km s$^{-1}$ 			&	 $\times$10$^3$ K cm$^{-3}$	&  $\times$10$^3$ K cm$^{-3}$	&  $\times$10$^2$ cm$^{-3}$  & $\times$10$^2$ cm$^{-3}$  & 	 \\
       &   observed  & in \hi  gas  & in \h2 gas &  \ciis &  \ci   &  \ciis & CO &  \\
\hline
BD +31 643	&	19.8	$\pm$	0.18	&	0.79	&	19.40	$\pm$	0.43	&	86 	    $\pm$	1.4	     &	---	&	12	$\pm$	0.2	&	---	& PDR (B5 V)		\\
HD 24534  	&	1.82	$\pm$	0.26	&	0.18	&	1.73	$\pm$	0.27	&	9.3	    $\pm$	1.4	      &	14.8	&	1.6	$\pm$	0.2	&	1.6	& 	\\	
HD 34078   	&	13.3	$\pm$	0.25	&	0.51	&	13.04	$\pm$	0.36	&	96 	    $\pm$	1.8	    &	---	&	13	$\pm$	0.2	&	---	& PDR (O9.5 Ve)		\\
HD 37021    &	101.5	$\pm$	0.22	&	1.58	&	100.71	$\pm$	0.82	&	400 	$\pm$	3.2	 &	6.8	&	56	$\pm$	0.5	&	---	& PDR $\theta^1$ Ori B (B3 V)\\		
HD 37061    &	118	    $\pm$	0.33	&	1.77	&	117.11	$\pm$	0.95	&	375 	$\pm$	3.0	   &  19.1	&	54	$\pm$	0.4	&	---	&PDR $\nu$ Ori (B0.5 V)\\		
HD 37903   	&	133.43	$\pm$	0.22	&	0.48	&	133.19	$\pm$	0.32	&	580	    $\pm$	1.4	&	40.7	&	85	$\pm$	0.2	&	$<$ 0.3	&PDR (B1.5 V)\\		
HD 62542   	&	3.23	$\pm$	0.21	&	0.28	&	3.09	$\pm$	0.25	&	31.1	$\pm$	2.1	    &	--- 	&	7.2	$\pm$	0.5	&	---	&PDR (B3 V)?		\\
HD 73882   	&	2.44	$\pm$	0.42	&	0.43	&	2.23	$\pm$	0.47	&	8.3	    $\pm$	1.7	&	--- 	&	1.6	$\pm$	0.3	&	---	&		\\
HD 94454   	&	1.06	$\pm$	0.2	   &	-0.02	&	1.07	$\pm$	0.20	&	7.2	    $\pm$	1.3	&	4.0	&	1.0	$\pm$	0.2	&	0.2 - 1.0	&		\\
HD 96675  	&	1.41	$\pm$	0.27	&	0.15	&	1.33	$\pm$	0.28	&	7.8	    $\pm$	1.6	&	--- 	&	1.3	$\pm$	0.3	&	0.8 - 4.0	&  		\\
HD 115071 	&	1.48	$\pm$	0.21	&	0.27	&	1.35	$\pm$	0.25	&	10.0	$\pm$	2.0	&	4.4	&	1.4	$\pm$	0.3	&	0.1 -0.8	&		\\
HD 144965	&	0.51	$\pm$	0.2	     &	0.11	&	0.46	$\pm$	0.21	&	2.9	     $\pm$	1.3	&	6.3	&	0.4	$\pm$	0.2	&	1.3 - 1.6	&	\\	
HD 147683	&	0.13	$\pm$	0.31	&	---	  &	---	                      &	---	               	    &	7.8	&	---	 	&	2.5	&		\\
HD 147888	&	2.54	$\pm$	0.21	&	1.69	&	1.69	$\pm$	0.87	&	27.8	$\pm$	12.0	&	9.6	&	6.3	$\pm$	2.7	&	4.0 - 5.0	&		\\
HD 152590	&	2.42	$\pm$	0.38	&	0.77	&	2.03	$\pm$	0.54	&	29.0	$\pm$	6.7   	&	5.1	&	4.5	$\pm$	1.0	&	0.3 - 1.3	& PDR (O7.5 V)\\		
HD 154368  	&	3.2	    $\pm$	0.48	&	0.33	&	3.04	$\pm$	0.51	&	10.2	$\pm$	1.6  	&	--- 	&	2.0	$\pm$	0.3	&	1.0 -4.0	&	\\	
HD 167971 	&	4.63	$\pm$	0.33	&	1.31	&	3.97	$\pm$	0.74	&	25.5	$\pm$	4.1	     &	--- 	&	4.0	$\pm$	0.6	&	---	& PDR (O8 Ib)		\\
HD 168076  	&	3.44	$\pm$	0.26	&	0.14	&	3.37	$\pm$	0.27	&	31.2	$\pm$	2.2	    &	--- 	&	4.6	$\pm$	0.3	&	---	&PDR (O4 V)		\\
HD 170740 	&	3.24	$\pm$	0.21	&	0.30	&	3.09	$\pm$	0.26	&	17.0	$\pm$	1.4	     &	--- 	&	2.4	$\pm$	0.2	&	---	&		\\
HD 185418 	&	1.06	$\pm$	0.29	&	0.51	&	0.80	$\pm$	0.39	&	5.4	    $\pm$	2.5	  &	2.6	&	0.5	$\pm$	0.2	&	0.1 - 0.4	&		\\
HD 192639	&	2.16	$\pm$	0.33	&	0.64	&	1.84	$\pm$	0.46	&	12.9	$\pm$	3.1	&	4.8	&	1.3	$\pm$	0.3	&	$<$ 0.8	&		\\
HD 200775	&	38.5	$\pm$	0.2	     &	0.16	&	38.42	$\pm$	0.22	&	140	    $\pm$	0.8	&	--- 	&	32	$\pm$	0.2	&	---	& PDR (NGC2023)	\\	
HD 203938  	&	2.73	$\pm$	0.25	&	1.00	&	2.23	$\pm$	0.56	&	8.9	    $\pm$	2.1	&	--- 	&	1.2	$\pm$	0.3	&	---	&		\\
HD 206267  	&	3.17	$\pm$	0.37	&	0.66	&	2.84	$\pm$	0.50	&	17.0	$\pm$	2.7	&	4.4	&	2.6	$\pm$	0.4	&	2.0 - 2.5	&		\\
HD 207198  	&	1.62	$\pm$	0.37	&	0.45	&	1.39	$\pm$	0.43	&	8.5	    $\pm$	2.5	&	4.3	&	1.3	$\pm$	0.4	&	0.3 -0.8	&		\\
HD 207538 	&	2.06	$\pm$	0.29	&	0.69	&	1.72	$\pm$	0.45	&	8.4 	$\pm$	2.1	&	--- 	&	1.2	$\pm$	0.3	&	---	&		\\
HD 210839  	&	1.12	$\pm$	0.27	&	0.47	&	0.89	$\pm$	0.36	&	5.0	    $\pm$	2.0	&	4.6	&	0.7	$\pm$	0.3	&	0.5	&		\\

\hline
\end{tabular}\\
%\vspace{0.3cm}
\end{center}
\vspace{-0.25cm}
Note: (1) This paper; (2) \cite{Jenkins2011}; (3) \cite{Goldsmith2013}.
\label{tab:Table2}
\end{table*}

%\pagebreak
\section{Discussion}
\label{sec:discussion}

\subsection{  Molecular Gas Pressure}

 To understand better the scatter in the \cii intensity versus \h2 column density in Figure 7 and the effect of assuming a value for the fractional  \c+ abundance $x$(C/\h2),  we consider a slightly different approach for displaying the data.  We can rewrite  Eq. A.8 as

\begin{equation}
\frac{{\rm I}({\rm CII})_{H_2}}{N_{17}(C^+)_{H_2}} = 0.51\times10^{-4} P_{th}  ,
\end{equation}
 where  $N$(\c+)$_{H_2}$, expressed in units of 10$^{17}$ cm$^{-2}$,  is the \c+ column density in the \h2 gas corresponding to the column density $N$(\h2).     Thus  the ratio of \cii intensity to \c+ column density is a direct measure of the thermal pressure.  For the {\it FUSE} targets with measured \h2 column densities we can use $N$(\c+)$_{H_2}$ = $N$(\h2)$x$(C/\h2).
Therefore to interpret the \cii intensities in terms of the measured $N$(\h2), it is necessary to assume a value of $x$(\c+/\h2) which introduces considerable ambiguity in the derived parameters such as pressure and volume density. There exist large variations  in $x$(C/H) values in the literature.  For example, \cite{Esteban2013} fit a value $\sim$  3.2$\times$10$^{-4}$ for the solar neighborhood from  deep echelle spectrophotometry of  Galactic \hii regions \citep[][]{Esteban2013,Garcia2007}.   Using the solar abundance of $x$(C/H) $\sim$  2.8$\times$10$^{-4}$ as a reference \cite{Jenkins2009} derives depleted ISM gas phase carbon abundance $\sim$ 1.8 to 2.2$\times$10$^{-4}$. \cite{Parvathi2012} measure the ISM gas and dust-phase carbon abundance along 15 Galactic lines of sight and find values of gas phase $x$(C/H) between 0.7 and  4.6$\times$10$^{-4}$.  As seen from Table 1 in our sample there are 13 targets that have  column densities of \h2 from {\it FUSE} and $N$(\c+) from UV absorption studies.     (We refer to the targets which have $N$(\c+) from UV absorption studies as UV--\c+ sample and those with \h2 absorption as {\it FUSE}--\h2 sample.)  We can then use these targets   to calibrate our data and estimate a more realistic value for x(\c+/\h2) applicable to all LOSs in our diffuse \h2 sample  using the following approach.

 (1) First, for the targets in the UV--\c+ sample, we use the fraction of \h2 along the line of sight, $f$(\h2) = 2$N$(\h2)/(2$N$(\h2)+$N$(\his)), derived from the {\it FUSE} \h2  and \hi column densities, to estimate the column density fraction of \c+ in \h2, $N$(\c+)$_{H_2}$ = $N$(\c+)$f$(\h2) where the total $N$(\c+) is listed in Table 1.

(2) We then correlate $N$(\c+)$_{H_2}$ and $N$(\h2) for these 13 targets.   The slope of a linear fit to these data corresponds to $x$(\c+/\h2).  We find, $x$(\c+/\h2) = (4.4$\pm$0.5)$\times$10$^{-4}$ which is within the range of depletion with respect to H  ($x$(C/H) $\sim$ 1.8 -- 2.2$\times$10$^{-4}$) derived by \citep{Jenkins2009} (note $x$(C/\h2)  = 2$x$(C/H)) and in our analysis it is obtained by using a sub-set of the {\it FUSE} targets themselves.

  In Figure 8 we plot the ratio $I$(\ciis)$_{H_2}$/$N$(\c+)$_{H_2}$ for each target LOS against its \cii intensity $I$(\ciis)$_{H_2}$. The data points  shown as crosses represent the targets in the  {\it FUSE}--\h2 sample for which the $N$(\c+)$_{H_2}$ are calculated using the \h2 column density and   x(\c+/\h2)= 4.4$\times$10$^{-4}$.   To further demonstrate the reliability of this value  for $x$(\c+/\h2) in Fig. 8 we plot $I$(\ciis)$_{H_2}$/$N$(\c+)$_{H_2}$ versus $I$(\ciis)$_{H_2}$    using a subset of 14   targets for which we have both $N$(\c+) data determined by STIS UV absorption studies and have HIFI \cii detections. The data points are shown as filled triangles representing the UV--\c+ sample for which   the $N$(\c+)$_{H_2}$  is obtained using $N$(\c+) and $f$(\h2) as discussed above.     It may be noted that this data set is not dependent on the assumed values  for $x$(C/\h2).  Thus the distribution of these data points (filled triangles) in $I$(\ciis)$_{H_2}$/$N$(\c+) -- $I$(\ciis)$_{H_2}$ space  provides a ``bench mark'' for the data plotted using the measured $N$(\h2) (crosses).   The thermal pressure shown along the Y-axis on the right side  corresponds to the value of the \cii intensity to \c+ column density ratio,  $I$(\ciis)$_{H_2}$/$N$(\c+), shown along the Y-axis on the left (see Eq. 6). The degree of  overlap between the two sets of data (the crosses and filled triangles) can be regarded as an independent validation of the assumed value for $x$(\c+/\h2) and our overall approach for estimating thermal pressure using the \cii intensities.

 In a $I$(\ciis)--$P_{th}$ representation, as shown in Figure 8, our {\it FUSE} \h2  sample points (crosses)  are distributed from the lower left (low pressure regime with low \cii intensity)  to upper right (high pressure regime with high \cii intensity). This relationship seems to indicate that our \h2 cloud sample represents a population of clouds in an evolutionary sequence.  Thus the \cii brightness  of the cloud is a good indicator of the evolutionary status of the cloud:  low brightness tracing formation stages while the high brightness ones the more evolved clouds in which increase in thermal pressure due to  increase in density, UV field and gravity become more important.  Alternatively, the \h2 clouds in our {\it FUSE}   sample could represent different environments, from low to high pressure regions.

\begin{figure}[h]
%\hspace{-0.5cm}
\centering
\includegraphics[scale=0.425, angle =-90]{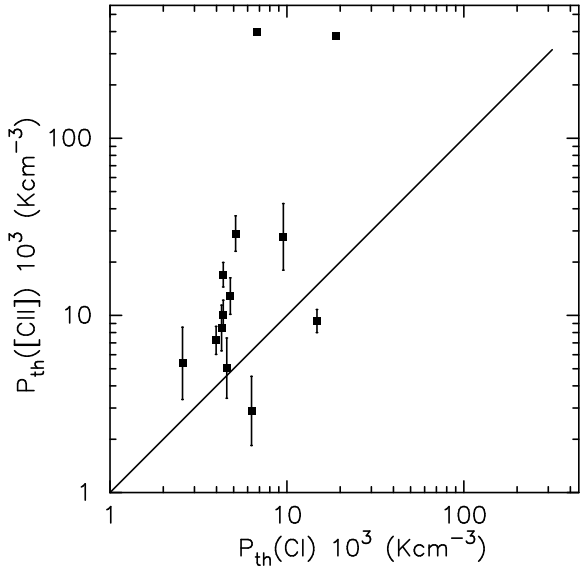}
%\hspace{0.5cm}
\caption{Comparison of the molecular gas pressure derived using the observed \cii intensities and \h2 column density  with those derived by JT2011. The solid line represents the two pressures being equal.
  }
\label{fig:f9}
\end{figure}

Observational data on the diffuse \h2 cloud properties such as mass, density, kinetic temperature, thermal and turbulent pressures, and their chemical make up are key to understanding the evolution of gas phases in the ISM.  \cite{Jenkins2011} have reviewed the status of the thermal gas pressure in diffuse \h2 clouds in the solar neighborhood.
Their measurements of thermal pressure  towards 89 target stars using the HST STIS absorption studies in the \ci fine structure lines provide  a large enough sample for comparison to our derivation of  thermal pressures using the HIFI \cii emission in concert with the \h2 UV absorption data.  There are a total of 15 target stars common to our {\it FUSE} \h2 sample and the JT2011 \ci sample, and we have \cii detections in 14 of them (\cii was not detected  in HD 147683 above a 5-$\sigma$  noise).  In Figure~\ref{fig:f9}  we compare the thermal pressures derived by JT2011 with those from \cii emission.

\begin{figure}[t]
%\hspace{-0.5cm}
\centering
\includegraphics[scale=0.385, angle=-90]{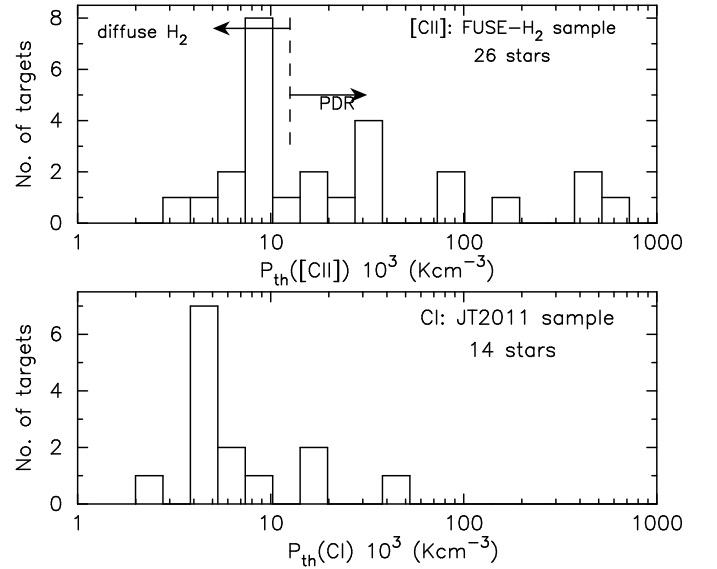}
%\hspace{0.5cm}
\caption{Number distribution of the molecular gas pressure (a) derived using the observed \cii intensities and \h2 column density , and (b)  \ci from JT2011. }
\label{fig:f10}
\end{figure}

As seen in Figure~\ref{fig:f9}  the thermal pressures derived from  $I$(\ciis) and  $N$(\h2) are typically higher than the corresponding \ci values in JT2011.  Part of this difference may be due to the sensitivity of \cii emission to the PDR and FUV environments, as discussed further in Section~\ref{sec:ExcessFarIR}.  To bring out this   difference  more clearly we compare in Figure~\ref{fig:f10} histograms of the number distribution of $P_{th}$ measured with \cii and \cis. The distribution maximum occurs for thermal pressures    $\sim$4000 to 5000 K cm$^{-3}$  in the \ci sample and   $\sim$7500 to 9000 K cm$^{-3}$  in the \cii sample. The peaks are similar enough that we can regard these sources as corresponding to the ISM pressure in diffuse \h2 gas along the LOS, while those with much higher thermal pressure, $P_{th} \geq$10$^4$ K cm$^{-3}$,  as  due to target-related PDR conditions. This bias is greatly enhanced by the likelihood for many of the target stars to occur in regions of high gas density especially,  because they are generally of early spectral type as identified in Table 2.   The 158 \microns \cii absorption measurements provides a more robust estimate of $N$(\c+)  \citep{Gerin2015}.  The thermal pressures derived from these absorption data are also in the range of 3800 to 22,000 K cm$^{-3}$ with a median value of 5900 K cm$^{-3}$.  These values are broadly consistent with our results from \cii emission spectra. However, there are several sources of uncertainties in all these comparisons.

% Figure 9

 In the {\it FUSE}--\h2 sample   among the total of 26 LOSs  for which we measured the thermal pressure using the \cii emission,  six have $P_{th}$ $>$ 50,000 K cm$^{-3}$ and   eight have  12,000 $<$ $P_{th}$ $<$ 30,000 K  cm$^{-3}$ while  12  out of 26,  have $P_{th}$ $\le$ 10000  K cm$^{-3}$.     In a sub-set of 14 LOSs common to both our \cii and JT2011 \ci targets,  seven LOSs   measured with \cii and eleven measured with \ci are in the low pressure regime with   $P_{th}$ $<$ 10,000 K cm$^{-3}$.   When we exclude the high pressure PDR gas ($P_{th}$ $>$ 10,000 K cm$^{-3}$) the  mean thermal pressure measured by \cii is   7700   K cm$^{-3}$,   which is     about 1.5 times higher than  5200 K cm$^{-3}$, measured by using  \cis.  The differences in distribution of thermal pressures measured with \cii and \cis, as  seen in the histograms in Figure~\ref{fig:f10}, indicates the differences in their intrinsic sensitivity to tracing \h2 gas in diffuse clouds.  Systematic effects  in the analysis of \cii emission and \ci absorption can lead to over-/under- estimating thermal pressures.   For example,  if low values for $x$(\c+/H) and $x$(\c+/\h2) are used  $P_{th}$ is overestimated  or the contribution to observed \cii intensity from \hi is underestimated. Furthermore, the uncertainties in $P_{th}$ listed in Table 2 are due to that in our estimate of the  \cii intensity in the \h2 gas which in turn is mostly due to that in the contribution from \hi (see Section 3.2.1 \& Eq. 3).  Thus $P_{th}$ is likely overestimated by underestimating the \hi contribution.  Indeed if we use the lower limit to $P_{th}$ as given by the uncertainties  in Table 2 in the {\it FUSE}--\h2 sample, we find 13 LOS with  $P_{th}$ $<$ 10,000 K cm$^{-3}$ with a mean pressure of 6100 K cm$^{-3}$ which is in better agreement with \ci results. We conclude that mean thermal pressure measured by \cii in diffuse \h2 gas is in the range of 6100 to 7700  K cm$^{-3}$.

\subsection{\h2 volume densities}

\cite{Goldsmith2013} constrained the  \h2 densities in diffuse molecular clouds using   an excitation  analysis of the UV absorption measurements of CO low lying rotational levels   up to J=3  towards many of the targets in our sample.  His procedure also uses the {\it FUSE} \h2 column densities that we use in our  \cii analysis.  He derives the \h2 density from  the excitation temperatures of the CO rotational levels seen in absorption, assuming a kinetic temperature typical of such clouds   (the results are not overly sensitive to $T_{kin}$ because the highest level observed, J=3, has an excitation energy of   33.2 K, much less than the typical $T_{kin}$  $\sim$ 70 K in the gas).  The CO excitation analysis uses the column densities of the J = 1--0, 2--1, and 3--2 transitions, where available, and the more levels detected the better the estimate of density.  In about 40\% of the sources excitation temperatures for two transitions are available, and in about 17\% all three were determined. When only one transition is detected only lower and upper limits on the densities can be determined.    In Table 2 we list the \h2 gas densities along the target LOSs derived using the \cii intensities and the respective \h2 column densities [Eq. 5].   For comparison with the \cii measurements, in Table 2 we show the  densities for 14 target LOSs derived from CO absorption data \citep{Goldsmith2013}.  In Figure 11 we show a scatter plot of the densities derived independently using the far-IR \cii and UV CO data.   The vertical bars for the densities measured with the CO data represents the range of densities constrained by the upper and lower limits given by \cite{Goldsmith2013}.  The  plot shows a reasonable agreement considering the different assumptions in the two approaches and that two outliers with high densities are in the high pressure PDR gas.
% Figure 11
\begin{figure}[h]
%\hspace{-0.5cm}
\centering
\includegraphics[scale=0.375, angle =-90]{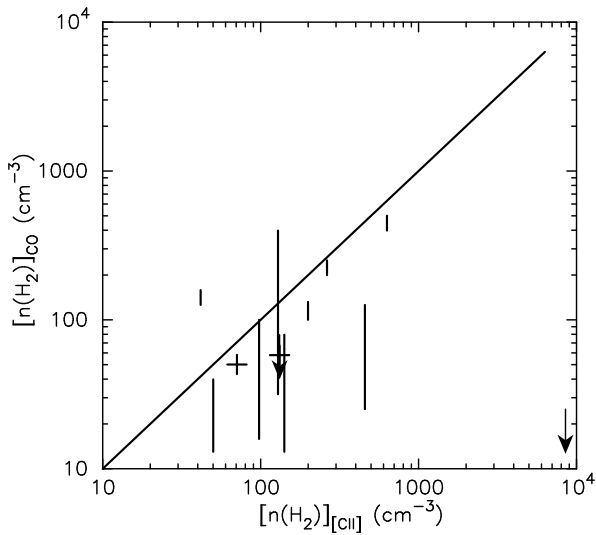}
%\hspace{0.5cm}
%\includegraphics[scale=0.4, angle = -90]{Fig_results_2.ps}
\caption{Comparison of the molecular gas densities derived using the observed \cii intensities and \h2 column density  with those derived by \cite{Goldsmith2013} from  CO excitation. The vertical error bars represent the upper and lower limits and the downward arrow the upper limit in Goldsmith's data. The solid line is defined by the two densities being equal.
  }
\label{fig:f11}
\end{figure}

\subsection{Excess far-IR \cii emission towards the target stars}
\label{sec:ExcessFarIR}

In a few cases  the proximity of the target star to the absorbing \h2 cloud becomes important as the target star plays the role of an exciting source creating a dense bright PDR environment resulting in larger \cii intensities.  In such cases the density and temperature structure within the \h2 cloud can vary significantly between the near-- and far--side to the observer.   These sources are readily distinguishable in UV by a rich spectrum of vibrationally excited \h2 observed by the STIS on HST  (e.g. \cite{Meyer2001}). In our  sample, in addition to HD37021 \& HD 37061, which lie in a strong UV and likely dense PDR environment in the Orion molecular cloud, there are several other target LOSs (BD +31 643, HD 34078HD 37903,  HD 147888, HD152590 and HD 200775) that have  \cii emission greatly enhanced by the presence of the PDRs. These are easily distinguished by their extremely high thermal pressure, $P_{th} >$ 15000 K cm$^{-3}$ (see Table 2).    The models for the PDR environment in HD 37903 \citep{Gnacinski2011}, in HD 147888 \citep{Gnacinski2013} show high densities and temperature structure as a  function of target distance on the near side of the cloud.  In the case of HD37903 the populations of the excited \h2 rovibrational states are   consistent with the   high density ($\sim$10$^4$ cm$^{-3}$) molecular gas located about 0.5 pc from the B1.5 V star.  BD +31 643 is a binary B5 V in IC 438, a young cluster (cf. \cite{Olofsson2012}). HD 200775 is the illuminating source of the reflection nebula NGC 7023 \citep{Federman1997}.  HD 34078 is a runaway star with a dense shell at the stellar wind/molecular
cloud interface that shows CH absorption and pronounced CO emission at the same V$_{LSR}$ velocities  as the \cii emission  \citep{Boisse2009}.
   Thus the presence of an enhanced UV field and ionization front  in these target environments  also enhance the \cii emission, thus yielding the high pressures we derived from \cii for these LOSs. The thermal pressures estimated from \cii intensities are underestimated, and represent likely  lower bound.   However, these targets require a detailed PDR modeling analysis  using the temperature and density structures \cite[cf.][]{Gnacinski2011}.

\subsection{  $(l-V)$ maps of \cii, \hi and CO emissions  }

The target HD 115071 is located towards the Crux spiral arm tangency and the ({\it l--V}) maps  (Fig.~\ref{fig:f3}a) show a snapshot of the \cii  and \hi  emissions, and the absence of CO at V$_{LSR}$$\sim$ -45 km s$^{-1}$ which bring out the differences among them  in spiral arm tangencies.   The  angular resolution of the the \hi 21 cm map (SGPS beam size $\sim$ 2.2\arcmin) is much lower than that of the CO or \cii maps, but has comparable velocity resolutions.   Therefore, we can compare the velocity structures.    Beyond the tangent velocity the \cii shows a relatively strong emission in contrast to that of  \hi  (with no CO)  consistent with the results reported by \cite{Velusamy2012,Velusamy2015}  that   the \cii emission  at this velocity  is tracing the compressed warm ionized medium (WIM) along this spiral arm tangency. The \cii emission in this feature delineating the compressed WIM is due to excitation by the electrons in contrast to the excitation by \h2 molecules in the features at V$_{LSR}$ $>$ -45 km s$^{-1}$ seen association with CO.  Similarly, in the {\it (l--V}) maps of HD 170740   we identify the enhanced \cii emission  with no CO,  seen at the highest velocities   near the tangent velocity at V$_{LSR}$ $\sim$ 80 \kmss as a likely WIM component.    In all three {\it (l-V)} maps the relatively strong \cii emission features,  which have associated CO, trace  the spiral arms along the LOSs:  the Scutum-Crux arm at V$_{LSR}$$\sim$ -45 to -30 km s$^{-1}$ (HD 115071), the  Scutum arm at  V$_{LSR}$ velocities near  $\sim$ 40 km s$^{-1}$ (HD 170740),   the Perseus spiral arm  at V$_{LSR}$  near  $\sim$ 20 km s$^{-1}$, and the Scutum and some of the Sagittarius-Carina arms at V$_{LSR}$ $\sim$ 30 km s$^{-1}$ (HD 168076).

\section{Summary}
\label{sec:summary}

The pressure in diffuse clouds is an important parameter in  understanding the processes that affect their origin and determine  their evolution. In this paper we have shown how \cii emission can be used to determine the thermal pressure and density in diffuse \h2 gas that is not readily traced in CO or other species.    The diffuse \h2 clouds can be identified by the presence of ionized carbon (\c+) emission (\ciis) without any CO emission.  Thus \cii provides an additional tool to complement \h2 and \ci studied by UV absorption  towards bright stars.

 In this paper, we present  \cii observations along  27 LOSs towards target stars of which 25 have  \textit{FUSE}  \h2 UV absorption data and demonstrate  how it could be used to estimate the thermal pressure in diffuse \h2 clouds.   We detect \cii 158 \microns emission features in 26 of these LOSs.   {For three LOSs which are close to the Galactic plane ${|\it b|} <$1\degs we also present longitude--velocity maps of \cii emission  observed by HIFI in on-the-fly (OTF) spectral line mapping.    We analyzed  and correlated the \cii velocity components detected along these LOSs in terms of the diffuse \h2 gas seen in the UV absorption data.}  We use the intensities of the \cii velocity components along with the column density of the diffuse \h2 gas as seen in the UV  to derive the thermal pressure and \h2 densities in the diffuse molecular clouds.      The distribution of thermal pressure in diffuse \h2 clouds shows a peak at  $\sim$ 7500 -- 9000 K cm$^{-3}$.  Our results provide a validation of the use of \cii line  intensity as a measure of  the \h2 gas in the diffuse \h2 molecular clouds.
 We illustrate  that the \cii intensity in \h2 clouds is a direct measure of the thermal pressure and  that it traces the low-- and high--pressure regimes in molecular cloud evolution.

Unlike the UV absorption studies which are limited to a pencil-like beam towards the direction of the target star, far-IR \cii line emission offers a much broader opportunity to study the diffuse \h2 gas over a cloud and throughout the Galaxy.  Spectral line mapping has the further advantage allowing us to study in detail their spatial and velocity structures. We present examples of such spectral line map data towards three target LOSs.  This use of \cii is particularly important in view of the new and enhanced opportunities from sub-orbital observations including  SOFIA.

%__________________________________________________________________

%----------------------
\begin{acknowledgements}
We thank the staffs of the ESA and NASA Herschel Science Centers for their help. We also thank an anonymous
referee for critical  comments and suggestions that improved
the analysis and discussion significantly. This work was performed at the Jet Propulsion Laboratory, California Institute of Technology, under contract with the National Aeronautics and Space Administration. %{\copyright}2016 California Institute of Technology. U.S. Government sponsorship acknowledged.
\end{acknowledgements}

\appendix

The thermal pressure in the interstellar medium is an important parameter for understanding interstellar cloud evolution but attempts to measure it in diffuse atomic and molecular clouds are generally restricted to lines of sight towards sources where absorption measurements of species such as neutral carbon and CO can be made in the UV and in the far-infrared  \ciis.  Here we discuss a method to derive the thermal pressure from the fine-structure emission line of \cii when the kinetic temperature is unknown.  The derivation is an extension of an earlier, but more restricted, solution suggested by  \cite{Langer2014_II} between intensity and pressure.  First we derive a solution in the low density limit applicable to a wide range of kinetic temperatures and then a more general solution, but valid over   a more restricted range of $T_{k}$.

\vspace{0.25cm}
\section{A. Low density solution}
\label{sec:PressureLow}

Here we derive an equation to relate $I$(\ciis) to the thermal pressure under some   realistic  restrictions on the gas kinetic temperature.  In the optically thin limit \cite[see Equation 25 in][]{Goldsmith2012}, the \cii intensity, $I$(\ciis) is given by
\begin{equation}
I({\rm [CII]}) = \int T_Adv = a_0 N_{3/2}({\rm C^+})\,\,{\rm (K\, km\,s^{-1})}  ,
\label{eqn:eq1}
\end{equation}
where
\begin{equation}
a_0 = \frac{hc^3A_{3/2,1/2}}{8\pi k \nu_{3/2,1/2}^2} =3.28\times 10^{-16} \,\,{\rm (K\, km\,s^{-1}\,cm^{-2} )}  ,
\label{eqn:eq2}
\end{equation}
\noindent  $A$= 2.3$\times$10$^{-6}$ s$^{-1}$, is the Einstein A-coefficient, the frequency  $\nu_{3/2,1/2}$=1.9005 THz, and  $N_{3/2}$ is the column density of the upper state in cm$^{-2}$. The units for $I$(\ciis) are  K \kms. For a uniform medium we can write $N_{3/2}$(C$^+$)=$n_{3/2}$(C$^+$)$L$, where $L$ is the path length in cm, and $n_{3/2}$(C$^+$) is the number density of ionized carbon in the upper $^2$P$_{3/2}$  state. For a two level system, in the optically thin regime, $n_{3/2}$(C$^+$) can be solved exactly, yielding
\begin{equation}
n_{3/2}({\rm C^+})=\frac{n_{tot}({\rm C^+})}{1+0.5[1+n_{cr}(X)/n(X)]e^{\Delta E/T_k}} ,
\label{eqn:eq3}
\end{equation}
\noindent where $n_{tot}$(C$^+$) is the total carbon ion density,  $n$(X) is the density for X, the dominant collision partner (H, H$_2$, or e), $n_{cr}$(X) is the critical density of X, $T_k$ is the kinetic temperature, and $\Delta E$ =91.21 K, is the excitation energy of the $^2$P$_{3/2}$ state in Kelvins. Combining these terms yields
\begin{equation}
I({\rm [CII]}) =\frac{a_0N({\rm C^+})}{1+0.5[1+n_{cr}(X)/n(X)]e^{\Delta E/T_k}}\,\,{\rm (K\, km\,s^{-1})} .
\label{eqn:eq4}
\end{equation}
\noindent In the low density limit where $n/n_{cr} <<$1, Equation~\ref{eqn:eq4} simplifies to,
\begin{equation}
\label{eqn:eq5}
I({\rm [CII]}) =2a_0N({\rm C^+})\frac{n(X)e^{-\Delta E/T_k}}{n_{cr}(X)}\,\,{\rm (K\, km\,s^{-1})} ,
\end{equation}
\noindent which can be rewritten in terms of the thermal pressure   $P_{th}^0$(X) =$n$(X)$T_k$, where X designates H or H$_2$, and the pressure is in units of  K cm$^{-3}$.
\begin{equation}
\label{eqn:eq6}
I({\rm [CII]}) =2a_0N({\rm C^+})\frac{P_{th}^0(X)}{n_{cr}({\rm X})}f(T_k)\,\,{\rm (K\, km\,s^{-1})} ,
\end{equation}
\noindent where we use $P^0$ to denote the thermal pressure in the low density limit.  The   term $f$($T_k$)=$T^{-1}_ke^{-\Delta E/T_k}$ is plotted in  Figure~\ref{fig:figA1}. As discussed in \cite{Langer2014_II} it  can be set to a constant value of  3.52$\times$10$^{-3}$ to within $\pm$15\% over the temperature range 45 to 200K, which is the typical temperature range for \cii emission. The resulting approximate solution for the pressure in the low density limit is
\begin{equation}
\label{eqn:eq7}
P_{th}^0({\rm X}) =4.14\times 10^{17} \frac{I({\rm [CII]})n_{cr}(X)}{N({\rm C^+})}\,\, ({\rm K\, cm^{-3}})  .
\end{equation}
\noindent The critical density for H$_2$ with an ortho-to-para ratio of one is $\sim$4.5$\times$10$^3$ cm$^{-3}$ at $T_k$= 100K \citep{Wiesenfeld2014} and is relatively insensitive to temperature, resulting in
\begin{equation}
\label{eqn:eq8}
P_{th}^0({\rm H_2}) =1.96\times 10^{4} \frac{I({\rm [CII]}}{N_{17}({\rm C^+})}\,\, ({\rm K\, cm^{-3}}) ,
\end{equation}
\noindent where we have expressed the column density, $N$(C$^+$), in natural units of 10$^{17}$ cm$^{-2}$, $N_{17}$(C$^+$). For atomic hydrogen regions $n_{cr}$(H) $\simeq$ 3.3$\times$10$^3$ cm$^{-3}$ for typical cloud temperatures, $T_k \sim$100K \citep{Goldsmith2012}, and in \hi clouds
\begin{equation}
\label{eqn:eq8b}
P_{th}^0({\rm H}) \simeq1.32\times 10^{4} \frac{I({\rm [CII]}}{N_{17}({\rm C^+})}\,\, ({\rm K\, cm^{-3}}) .
\end{equation}
% Figure A1
% Figure A1
\setcounter{figure}{0}
\begin{figure}[t]
\makeatletter
%\vspace{-1.0cm}
\renewcommand{\thefigure}{A.1}
\centering
            \includegraphics[width=7cm]{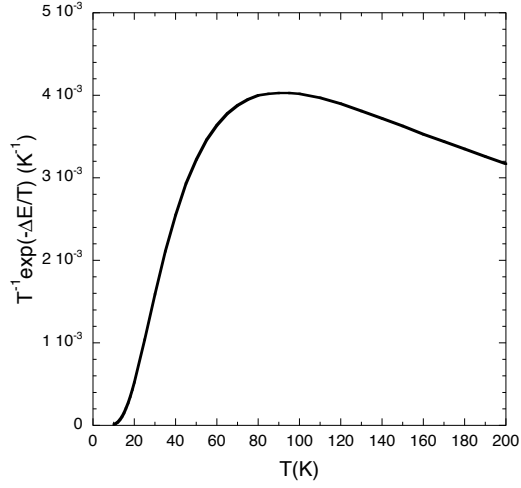}
        \caption{Dependence of $T^{-1}$exp(-91.21/T), which is proportional to the relative excitation rate to the $^2$P$_{3/2}$ level of C$^+$, on the kinetic  temperature. This term in the excitation rate varies by only ±15\% about its mean value of 3.52$\times$10$^{-3}$ over the range 45K $\leq T_{kin} \leq$ 200K, and by only ±20\% from 40K to 200K.}
         \label{fig:figA1}
 \end{figure}
\section{B. General solution}
\label{sec:PressureGeneral}
It is possible to generalize the solution for the thermal pressure to all densities, i.e. arbitrary $n$(X)$/n_{cr}$(X), but under a slightly more restrictive temperature range. Fortunately this range, $T_k \sim$ 65 to 125 K, is appropriate for most   \cii emission arising from atomic and molecular hydrogen regions. We start by rewriting Equation~\ref{eqn:eq4}  in terms of the thermal pressure, $P_{th}$, for arbitrary $n$(X)/$n_{cr}$(X), which gives
 \begin{equation}
I({\rm [CII]}) =\frac{a_0 N({\rm C^+})}{1+0.5e^{\Delta E/T_k} +[0.5n_{cr}(X)]/[P_{th}({\rm X})f(T)]}\,%\,{\rm (K\, km\,s^{-1})} .
\label{eqn:eq9}
\end{equation}
\noindent Solving for $P_{th}$ yields
 \begin{equation}
 P_{th}({\rm X})=\frac{0.5n_{cr}({\rm X})I([{\rm CII]})}{a_0f(T)N({\rm C^+})}\frac{1}{1-[b(T)I([{\rm CII}])]/[a_0 N({\rm C^+})]} ,
\label{eqn:eq10}
\end{equation}
\noindent where $b(T)$=1+0.5e$^{\Delta E/T_k}$.  Note that the first term in Equation~\ref{eqn:eq10} is just the low density solution, $P_{th}^0$(X), given in Equation~\ref{eqn:eq7}, so that
 \begin{equation}
% \frac{p_{th}}{p_0} =\frac{1}{1-(b(T)I({\rm CII})/(a_0 N({\rm C^+}))}.
 \frac{P_{th}}{P_{th}^0} =\Big[1-\frac{b(T)I([{\rm CII}])}{a_0 N_{17}({\rm C^+})}\Big]^{-1} .
\label{eqn:eq11}
\end{equation}
We see from Equation~\ref{eqn:eq11} that $P_{th}^0$ underestimates the true thermal pressure for large values of  $I$(\ciis)/$N_{17}$(C$^+$), which are characteristic of higher density regions.  In Figure~\ref{fig:figA2} we plot $P_{th}/P^0_{th}$ in Equation~\ref{eqn:eq11} versus $I$(\ciis)/$N_{17}$(C$^+$) for a range of relevant kinetic temperatures. It is clear that the low density solution is excellent for $I$(\ciis)/$N_{17}$(C$^+$)$\leq$3, and reasonable up to a ratio of 5, but beyond that the solution to the pressure is sensitive to $T_k$.

If we do not know the kinetic temperature we see that  from Figure~\ref{fig:figA2}   over the important temperature range, 65 K to 125 K,   we can set $b$(T) to a constant value and get a reasonable estimate of $P_{th}$.  We find that fixing  $b$(T) = 2.70 fits $P_{th}/P^0_{th}$ to within $\pm$25\% over $T_k$ = 60 to 125 K, and to better than $\pm$20\% over 65 to $\sim$110 K.  This substitution, when used with a fixed value for $f(T_k)$, yields an approximate expression for $P_{th}$(X) independent of density,
\begin{equation}
% \frac{p_{th}}{p_0} =\frac{1}{1-7.94\times 10^{-2}I({\rm CII})/N({\rm C^+})}.
P_{th} =\Big[1-8.23\times10^{-2}\frac{I([{\rm CII}])}{N_{17}({\rm C^+})}\Big]^{-1} {P_{th}^0} ,
\label{eqn:eq12}
\end{equation}
\noindent which has a solution for $I$(\ciis)/$N_{17}$(C$^+$)$\lesssim$12.1. We plot the ratio $P_{th}/P_{th}^0$ from Equation B.4 in Figure~\ref{fig:figA2} (dashed line) and it can be seen to provide a reasonable solution for the temperature range 65 to 125 K, when $T_k$ is not well known.  We do not recommend estimating the true thermal pressure for very large values of the ratio $I$(\ciis)$/N_{17}$(C$^+$)$\geq$5, unless the kinetic temperature is reasonably  well known.
% Figure A2
\setcounter{figure}{0}
\begin{figure}[t]
\makeatletter
%\vspace{-1.0cm}
\renewcommand{\thefigure}{B.1}
 \centering
            \includegraphics[width=7cm]{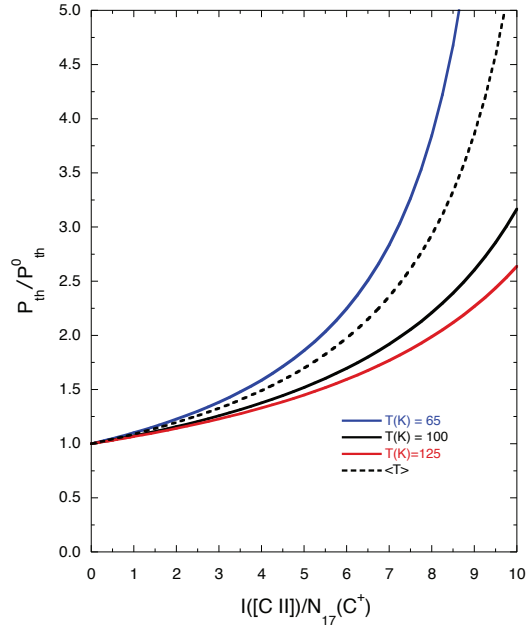}
        \caption{The ratio of the thermal pressure to the low density solution as a function of $I$(\ciis)/$N_{17}$(C$^+$) for a range of $T_k$ (solid lines). The approximate solution when the kinetic temperature is not known and a fixed value is adopted for $b(T_k)$ is shown by the dashed line (Eq. B.4)}
        \label{fig:figA2}
 \end{figure}
To test how well Equation B.4 reproduces the pressure given the ratio $I$(\ciis)/$N$(C$^+$) we compared the results to a radiative transfer model assuming a Large Velocity Gradient (LVG).  We  used the RADEX radiative transfer code  \citep{vanderTak2007} and the LAMDA data base \citep{Schoier2005} to calculate $I$(\ciis) as a function of $P$ = $nT$ and column density, $N$(C$^+$).  In contrast to our optically thin solution, the RADEX code includes the effects of \cii opacity, $\tau$(\ciis).  In the RADEX code we fixed $T_k$ = 100 K, the linewidth $\Delta V$ = 5 and 10 \kms, and the column density to $N$(C$^+$) = 5$\times$10$^{17}$ and 2$\times$10$^{18}$ cm$^{-2}$  to represent a typical range in our data set.  We then varied the density $n$(H$_2$) and calculate $P$=$nT$ and $I$(\ciis). We found good agreement within 15 -- 20\% between the approximate and RADEX solutions for $P_{th}$.
%__________________________________________________________________

%%%%%% REFERENCES %%%%%%%%%%%%%%%%%%%%%%%%%%
%% the following two command lines are needed for using bibtex
%\bibliographystyle{aa}
\bibliographystyle{apj}
\bibliographystyle{aasjournal.bst}
\bibliography{aa_CII_spiral_refs_v3}
\end{document}